\documentclass[12pt]{iopart}
\usepackage{epsfig,bm}
\usepackage{graphicx,cite}
\usepackage{amssymb,psfrag,textcomp}
\usepackage{amsthm,upgreek}

\newcommand{\bracket}[1]{\left\langle #1\right\rangle}

\newcommand{\be}{\begin{equation}}
\newcommand{\ee}{\end{equation}}
\newcommand{\bd}{\begin{displaymath}}
\newcommand{\ed}{\end{displaymath}}

\newcommand{\openone}{\mathcal{I}}

\newcommand{\talpha}{\tilde{\alpha}}

\def\Xint#1{\mathchoice
{\XXint\displaystyle\textstyle{#1}}%
{\XXint\textstyle\scriptstyle{#1}}%
{\XXint\scriptstyle
      \scriptscriptstyle{#1}}%
{\XXint\scriptscriptstyle
      \scriptscriptstyle{#1}}%
\!\int}
\def\XXint#1#2#3{{
   \setbox0=\hbox{$#1{#2#3}{\int}$}
\vcenter{\hbox{$#2#3$}}\kern-.5\wd0}}
\def\dashint{\Xint-}

\begin{document}
\title{Large deviations of the shifted index number in the Gaussian ensemble}
\author{Isaac P\'erez Castillo}
\address{Departamento de Sistemas Complejos, Instituto de F\'isica, UNAM, P.O. Box 20-364, 01000 M\'exico, D.F., M\'exico}
\begin{abstract}
We show that, using the Coulomb fluid approach, we are able to derive a rate function $\Psi(c,x)$ of two variables that captures: (i) the large deviations of bulk eigenvalues; (ii) the large deviations of extreme eigenvalues (both left and right large deviations); (iii) the statistics of the fraction $c$ of eigenvalues to the left of a position $x$. Thus, $\Psi(c,x)$  explains the full order statistics of the eigenvalues of large random Gaussian matrices  as well as the statistics of the shifted index number. All our analytical findings are thoroughly compared with Monte Carlo simulations, obtaining excellent agreement. A summary of preliminary results was already presented in \cite{Perez2014b} in the context of one-dimensional trapped spinless fermions in a harmonic potential.
\end{abstract}
\maketitle
\section{Introduction}
Recently, ensembles of random matrices have become the perfect mathematical laboratory  to answer a question which is ubiquitous in many branches of sciences: how  universal statistical properties of weakly correlated random variables are affected when correlations become strong and how these properties emerge in systems of interest.\\
Most of the work in random matrix theory has thus focused to see how  limiting distributions of extreme value statistics\cite{Fisher1928,Gnedenko1943,Galambos1978,Haan2006,Gumbel1958,Weibull1951} change in the presence of strong correlations \cite{Dean2006,Dean2008,Marino2014,Majumdar2012,Vivo2007,Vivo2008,Majumdar2009, Fachi2008, Katzav2010, Nadal2010,Nadal2011,Fyodorov2012,Majumdar2013,Majumdar2014, Perez2014, Ramli2012,Rourke2010,Gustavsson2005,Johansson2000}. However, in this framework the study of the so-called order statistics,  that is the  statistics of the $k$-th eigenvalue, remains essentially mostly untapped and only few works  are available  \cite{Gustavsson2005,Rourke2010}, but even those  have only explored the typical behaviour.\\
In this work, we generalise the results presented in \cite{Gustavsson2005,Rourke2010} to describe typical as well as atypical fluctuations in the asymptotic limit of very large matrices. We derive, in turn, a unifying way to obtain large deviations to the left and to the right of the typical value of an extreme eigenvalue.  In the past,  large deviations to the left and to the right of, for instance, the largest eigenvalue had to be analysed separately, using the Coulomb fluid picture to obtain the left rate function (see, for instance, \cite{Dean2006,Dean2008}), and other methodologies (e.g. energetic balance arguments \cite{Majumdar2009}, formal mathematical approaches \cite{Johansson2000}, other rather elaborate mathematical approaches as loop theory) to describe the deviations to the right. Surprisingly, as we will illustrate, these two statistics for bulk and extreme eigenvalues are captured by a single rate function $\Psi(c,x)$ depending on two parameters.\\
Part of these results already appeared in \cite{Perez2014b} in the context of full order statistics of a system of  one-dimensional spinless fermions trapped on a external harmonic potential. Here we present all details of the mathematical derivations and compare our findings thoroughly with Monte Carlo simulations.\\
This paper is organised as follows: in Sect \ref{sec:md} we introduce the various definitions that will be used throughout the paper and we also point out the connection of the full order statistics with the statistics of the shifted index number; in Sect \ref{sec:cfa} we describe the mathematical derivations using the Coulomb fluid picture, exposing as well the various ways to evaluate the free energy associated to the Coulomb fluid, how to derive exact expressions of the constrained spectral density resulting from fixing a fraction of eigenvalues to the left of  $x$, and how to obtain an exact expression of the rate function using complete and incomplete elliptic integrals; in Sect. \ref{sec:tcd} the tail cumulative distribution for the shifted index number is presented together with its connection to the statistics of the $k$-th eigenvalue; in Sect. \ref{sec:ldf} we focus on recovering  large deviations of  extreme eigenvalues while in Sect. \ref{sec:tfa} we deal with the typical fluctuations of  bulk eigenvalues; in Sect. \ref{sec:mcs} we describe the various Monte Carlo simulations performed in this work. Finally,  Sect. \ref{sec:cfw} contains  concluding remarks and describes  possibly interesting future research lines.
\section{Model definitions}
\label{sec:md}
We start with some definitions. Given a random variable $X$ taking values $x\in\Omega$, we denote as $F_X(x)={\rm Prob}[X\leq x]$ its cumulative density function (CDF) and $\overline{F}_X(x)=1-F_{X}(x)$ its tail CDF. Similarly, we denote the probability density function (PDF) as $f_{X}(x)={\rm Prob}[X=x]$.\\
We are interested in studying the statistical properties related to the joint Probability Density Function (jPDF) of eigenvalues $\bm{y}=(y_1,\ldots,y_N)$ of the Gaussian ensemble
\begin{equation}
P(\bm{y})=\frac{1}{A_0}e^{-\frac{\beta}{2}\sum_{i=1}^{N}y^{2}_i}\prod_{i<j}|y_i-y_j|^{\beta}\,,\nonumber
\end{equation}
where $\beta$ is Dyson's index, $A_{0}$ is a normalising factor for $P(\bm{y})$ such that $-\infty<y_i<\infty$ for $i=1,\ldots,N$. In particular, we focus on what we call the Shifted Index Number (SIN), as the random variable $\mathcal{N}_x=\sum_{i=1}^N\Theta(x-y_i)$, which counts  the number of eigenvalues to the left of $x$ and can take values from the set $n_{x}\in\{0,\ldots,N\}$.  Its PDF is
\begin{equation*}
f_{\mathcal{N}_x}(n_x)=\int_{-\infty}^{\infty}d\bm{y}P(\bm{y})\delta\left(n_x-\sum_{i=1}^N\Theta(x-y_i)\right)\,.
\end{equation*}
Even though the SIN takes discrete values we use Dirac deltas instead of Kronecker deltas as this distinction becomes unimportant in the limit of very large matrices. Its corresponding tail CDF is obviously given as\footnote{The upper limit of the integral in the definition of $\overline{F}_{\mathcal{N}_x}(n_x)$  should  actually be $N$, but since we are interested in the asymptotic limit of very large matrices, this has been replaced by infinite.}
\begin{equation*}
\overline{F}_{\mathcal{N}_x}(n_x)=\int_{n_x }^{\infty} d y f_{\mathcal{N}_x}(y)\,.
\end{equation*} 
Next, we make the following remark: the probability that the $k$-th eigenvalue $y_k$ is smaller than $x$ is precisely the probability that at least $\mathcal{N}_x$ is greater than $k$, or in other words $\overline{F}_{\mathcal{N}_x}(k)=F_{y_k}(x)$. Alternatively, we have that $F_{\mathcal{N}_x}(k)=\overline{F}_{y_k}(x)$. Thus, by studying $F_{\mathcal{N}_x}$ (or its PDF $f_{\mathcal{N}_x}$) we have access not only to the statistical properties of SIN, but also to those of the $k$-th eigenvalue. This modest generalisation of the index distribution problem \cite{Majumdar2009b}, when analysed using the Coulomb fluid method,  is able to provide in the limit of large matrices all statistical properties of bulk and extreme eigenvalues. 

\section{Coulomb fluid approach}
\label{sec:cfa}
To obtain an expression of $f_{\mathcal{N}_x}$ for very large matrices we use the Coulomb fluid approach (see, for instance, \cite{Majumdar2009b} for the treatment of the standard index number). We  first write the jPDF of eigenvalues as $P(\bm{\lambda})=\frac{1}{Z_{0}}e^{-\beta N^2 E(\bm{\lambda})}$ with $E(\bm{\lambda})=\frac{1}{2N}\sum_{i=1}^N\lambda_{i}^2-\frac{1}{2 N^2}\sum_{i\neq j}\log|\lambda_i-\lambda_j|$. Here we have scaled the eigenvalues as $y_i=\sqrt{N}\lambda_i$ and have also scaled the position of the barrier $x$ accordingly, $x\to\sqrt{N} x$. We also introduce the intensive random variable $\mathcal{C}_x=\mathcal{N}_x/N$, which takes values $c\in[0,1]$ for large $N$. This allows us to write $\varrho(c)\equiv f_{C_{x}}(c)={\rm Prob}[\mathcal{C}_x=c]$ as
\begin{equation*}
\varrho(c)=\frac{1}{A_0}\int_{-\infty}^{\infty}d\bm{\lambda}e^{-\beta N^2 E[\rho(\lambda;\bm{\lambda})] }\delta\left( c-\int \rho(\lambda;\bm{\lambda})\Theta(x-\lambda)\right)\,,\nonumber
\end{equation*}
with $\rho(\lambda;\bm{\lambda})=(1/N)\sum_{i=1}^N\delta(\lambda-\lambda_i)$ and
 \begin{equation*}
\hspace{-1cm}E[\rho(\lambda;\bm{\lambda})]=-\frac{1}{2}\int d\lambda d\lambda' \rho(\lambda)\rho(\lambda')\log|\lambda-\lambda'|+\frac{1}{2}\int d\lambda\rho(\lambda)\lambda^2+\mathcal{O}(N^{-1})\,.
\end{equation*}
Next, we introduce a functional Dirac delta and express it in its Fourier representation, viz.
\begin{eqnarray}
\varrho(c)&&=\frac{1}{Z_0}\int D[\rho,\gamma] e^{-\beta N^2 E[\rho(\lambda)]+iN\int d\lambda \gamma(\lambda)\rho(\lambda)+N\log \int_{-\infty}^{\infty}d\lambda  e^{-i\gamma(\lambda)}}\nonumber\\
&&\times\delta\left( c-\int d\lambda\rho(\lambda)\Theta(x-\lambda)\right)\,,\label{eq:fu}
\end{eqnarray}
 with $Z_0$ is a normalisation constant. Henceforth whichever constant appears during the subsequent derivations will be absorbed into the constant $Z_0$. Using variational calculus on the path $\gamma(\lambda)$ yields the following saddle-point equation
\begin{equation}
\rho(\lambda)=\frac{e^{-i\gamma(\lambda)}}{\int_{-\infty}^{\infty}d\lambda'  e^{-i\gamma(\lambda')}}\,,\nonumber
\end{equation}
which simply states that $\rho(\lambda)$ is normalised. Plugging  this solution  back into \eref{eq:fu} and after some standard manipulations,  one obtains the PDF of the fraction $c$ of eigenvalues to the left of $x$ as:
\begin{eqnarray}
\hspace{-1cm}\varrho(c)&=&\frac{1}{Z_0}\int D[\rho,A_1,A_2] e^{-\frac{\beta}{2} N^2S[\rho,A_1,A_2] }\,,\\
\hspace{-1cm} S[\rho,A_1,A_2]&=&\int d\lambda\rho(\lambda)\lambda^2-\int d\lambda d\lambda' \rho(\lambda)\rho(\lambda')\log|\lambda-\lambda'|\nonumber\\
\hspace{-1cm}&+&A_1\left(\int_{-\infty}^{\infty}d\lambda \rho(\lambda)\Theta(x-\lambda)-c\right)+A_2\left(\int_{-\infty}^{\infty} d\lambda \rho(\lambda)-1\right)\,,
\label{eq:action}
\end{eqnarray}
where $D[\rho,A_1,A_2]$ means integrating over the set of functions $\rho(\lambda)$ and variables $A_1$ and $A_2$. For large $N$, $\varrho(c)$ can be evaluated by the saddle-point method yielding $\varrho(c)=\exp\left[-\beta N^2\Psi(c,x)\right]$, expressed in terms of the rate function
\begin{equation}
 \Psi(c,x)=\frac{1}{2}\left(S_{0}(c,x)-\Omega_0\right)\,,
\label{eq:rf}
\end{equation}
with $\Omega_0=\frac{3+2\log(2)}{4}$. Here $S_{0}(c,x)$ refers to the action \eref{eq:action} evaluated at the saddle point obtained by seeking stationarity  with respect to $\rho$, $A_1$, and $A_2$, while $\Omega_0$ refers to the action related to the normalising constant $Z_0$.  The variation of the action with respect to those parameters, that is
\begin{equation*}
\frac{\delta S[\rho,A_1,A_2]}{\delta \rho(\lambda)}=\frac{\partial S[\rho,A_1,A_2]}{\partial A_1}=\frac{\partial S[\rho,A_1,A_2]}{\partial A_2}=0\,,
\end{equation*}
provides the saddle-point equations:
\numparts
\begin{eqnarray}
\lambda^2&+&A_1\Theta(x-\lambda)+A_2=2\int d\lambda' \rho(\lambda')\log|\lambda-\lambda'|\label{eq:spe1}\\
1&=&\int_{-\infty}^{\infty} d\lambda \rho(\lambda)\,, \label{eq:spe2}\\
c&=&\int_{\infty}^{\infty}d\lambda \rho(\lambda)\Theta(x-\lambda)\,.\label{eq:spe3}
\end{eqnarray}
\endnumparts
The set of equations \eref{eq:spe1}, \eref{eq:spe2} and \eref{eq:spe3}  reflect that  we are seeking for the stationary solution in the thermodynamic limit of the charge density of a  two-dimensional Coulomb fluid contrained into a one-dimensional line, with external harmonic potential in which a fraction $c$ of charges (read eigenvalues) must be to the left of $x$. If the constraint \eref{eq:spe3} is  absent  the solution of the equations \eref{eq:spe1} and \eref{eq:spe2}   is the celebrated Wigner's semi-circle law $\rho_{{\rm sc}}(\lambda)=\frac{1}{\pi}\sqrt{2-\lambda^2}$.  Alternatively, one could also obtain the semi-circle law even in the presence of such constraint, if the constraint turns out to be  irrelevant.  This corresponds to the case  when,  given $x$, the fraction $c$ is  precisely the fraction of eigenvalues to the left of $x$ in Wigner's law.  Let us denote this value  $c_{\star}(x)$. Then, with a modest amount of foresight, we can anticipate two distinct regimes giving rise to deformations of Wigner's semicircle law: either $c>c_{\star}(x)$ or $c<c_{\star}(x)$.\\
As the main goal is to formulate an expression for the rate function $\Psi(c,x)$, we proceed as follows: i) we first rewrite  the action in terms of the second moment of the constrained density and two constants; ii) we then solve  the saddle-point equations \eref{eq:spe1}, \eref{eq:spe2} and \eref{eq:spe2} to obtain exact expressions for the deformed semicircle law; iii) we finally work out formulas for its second moment and those two constants.
\subsection{Rewriting the action $S_0(c,x)$ at the saddle-point }
Multiplying the  saddle-point equation  \eref{eq:spe1} by $\rho(\lambda)$ and integrating over $\lambda$ gives
\begin{equation*}
\int d\lambda d\lambda' \rho(\lambda)\rho(\lambda')\log|\lambda-\lambda'|= \frac{1}{2}\int d\lambda\rho(\lambda)\lambda^2+\frac{A_1}{2}c+\frac{A_2}{2}\,.
\end{equation*}
We next use this result to replace the double integral appearing in the action \eref{eq:action} in terms of one single integral and two constants $A_1$ and $A_2$, \textit{viz.}
\begin{equation}
S_0(c,x)=\frac{1}{2}\int d\lambda\rho(\lambda)\lambda^2-\frac{A_1}{2}c-\frac{A_2}{2}\,.
\label{eq:action2}
\end{equation}
Alternatively, and as it was reported in \cite{Perez2014b},  recall that the variation with respect to external parameters at the saddle point enters only through its explicit derivatives, that is
\begin{eqnarray*}
\frac{\partial S_0(c,x)}{\partial c}&=&\frac{\partial S[\rho,A_1,A_2]}{\partial c}+\frac{\delta S[\rho,A_1,A_2]}{\delta \rho}\frac{\partial \rho}{\partial c}\\
&&+\frac{\partial S[\rho,A_1,A_2]}{\partial A_1}\frac{\partial A_1}{\partial c}+\frac{\partial S[\rho,A_1,A_2]}{\partial A_2}\frac{\partial A_2}{\partial c}\\
&&=\frac{\partial S[\rho,A_1,A_2]}{\partial c}\,,
\end{eqnarray*}
which yields
\begin{equation*}
\frac{\partial S_0(c,x)}{\partial c}=-A_1(c,x)
\end{equation*}
and, therefore, integrating over $c$ we have
\begin{equation*}
S_{0}(c,x)-\Omega_0=\int^{c}_{c_\star(x)} dc' \frac{\partial S_0(c',x)}{\partial c'}=-\int^{c}_{c_\star(x)} dc'A_1(c',x)\,.
\end{equation*}
Here,  we have used that $S_0(c_\star(x),x)=\Omega_0$ or, alternatively, $\Psi(c_{\star}(x),x)=0$. From expression \eref{eq:action2} we see that, to evaluate the action at the saddle point, we must arrive at formulas for the second moment of $\rho(\lambda)$ and constants $A_1$ and $A_2$. 
\subsection{Solving the saddle-point equations. A complete picture of the constrained spectral density}
\begin{figure*}[t]
\includegraphics[width=4.5cm,height=5cm]{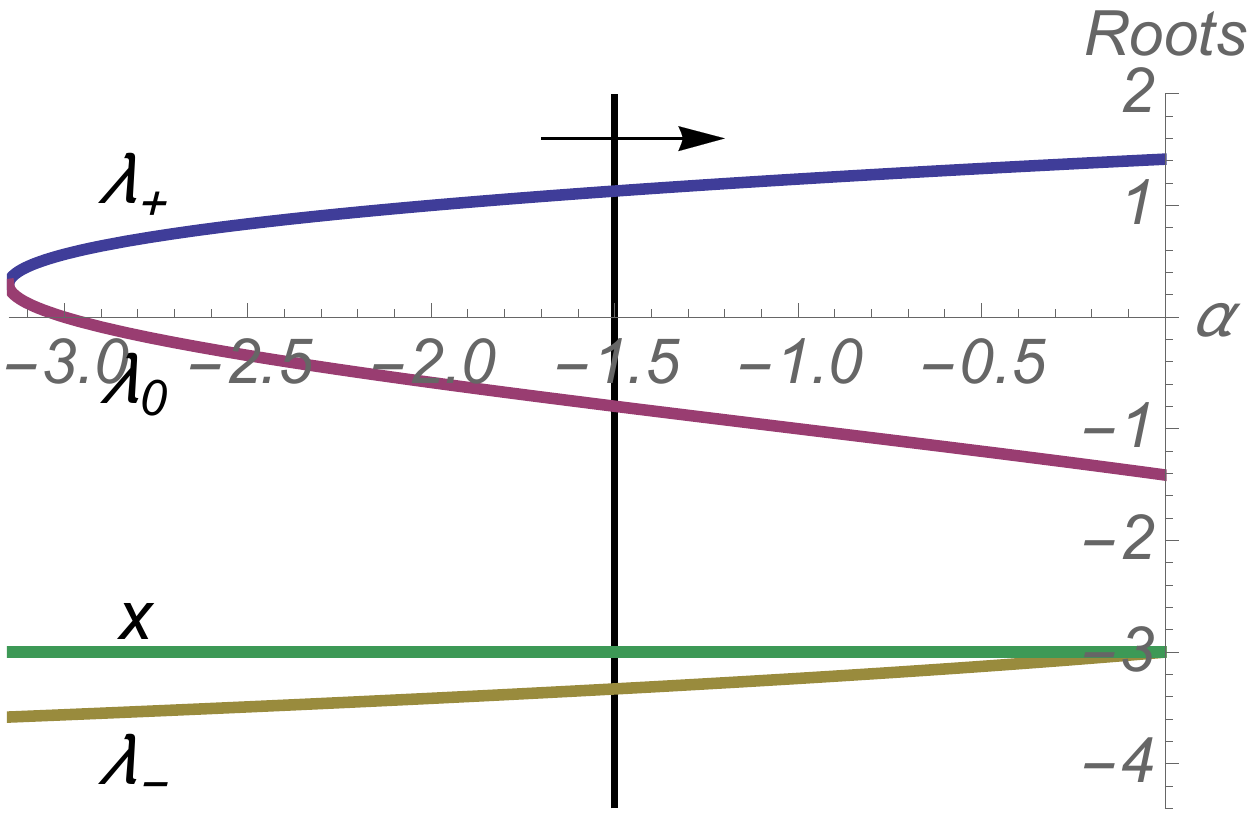}\quad\quad
\includegraphics[width=4.5cm,height=5cm]{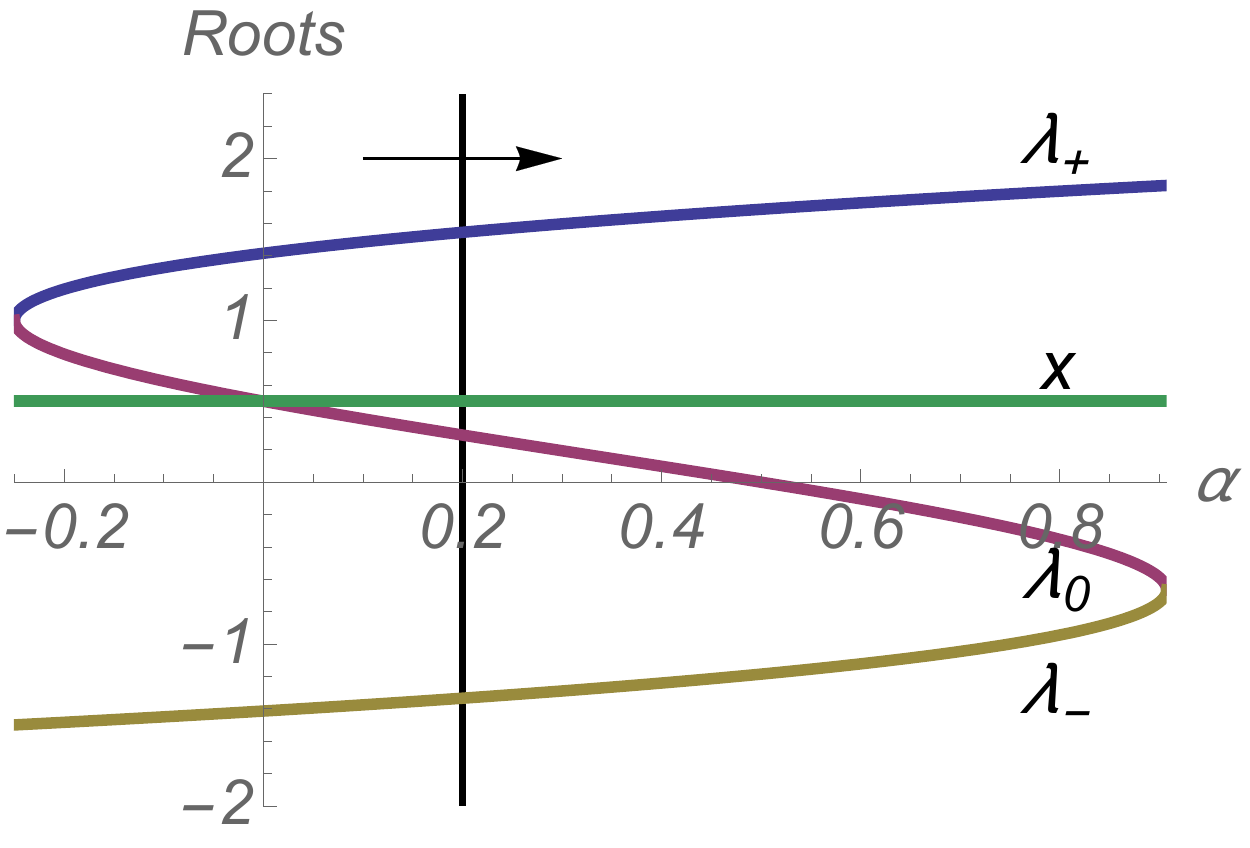}\quad\quad
\includegraphics[width=4.5cm,height=5cm]{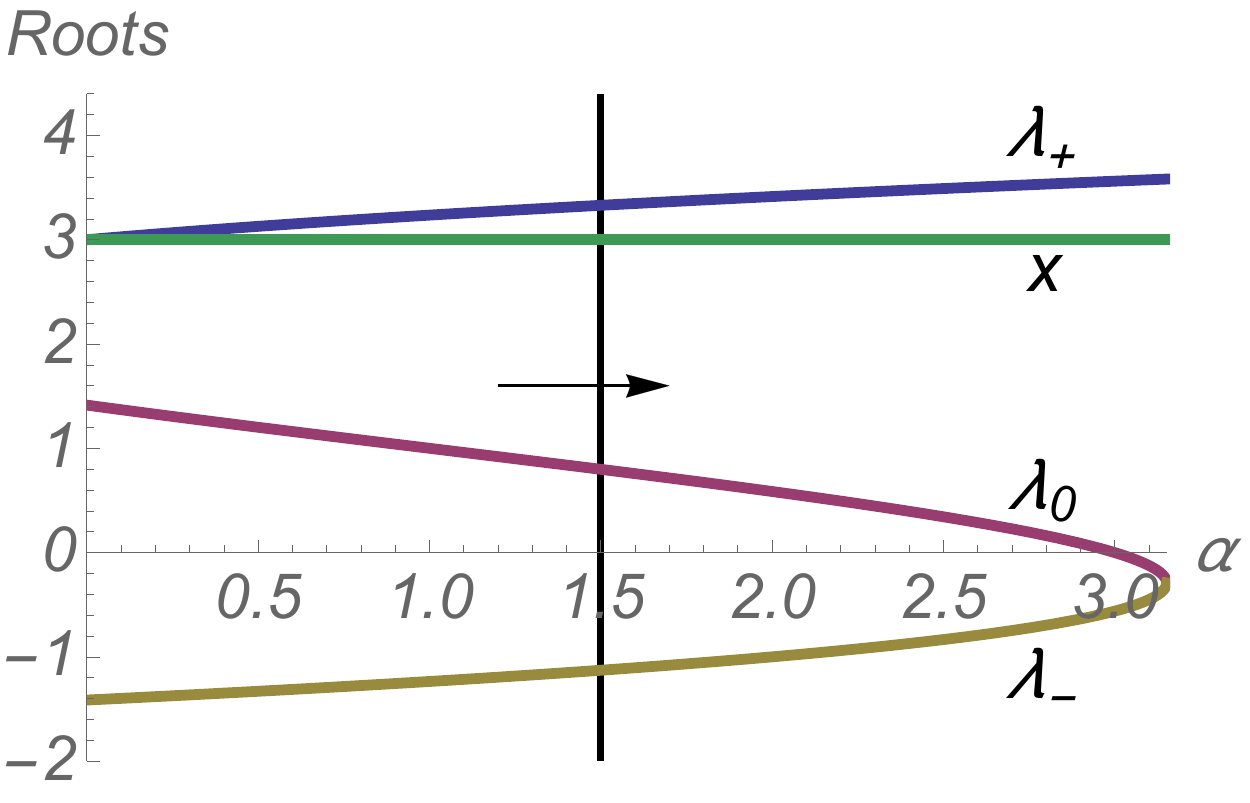}
\caption{Plot of the three roots as a function of $\alpha$ for fixed $x=-3,1/2,3$ (from left to right). Explanation can be found in the text.}
\label{roots}
\end{figure*}

\begin{figure}[h]
\centering
\includegraphics[width=10cm,height=7cm]{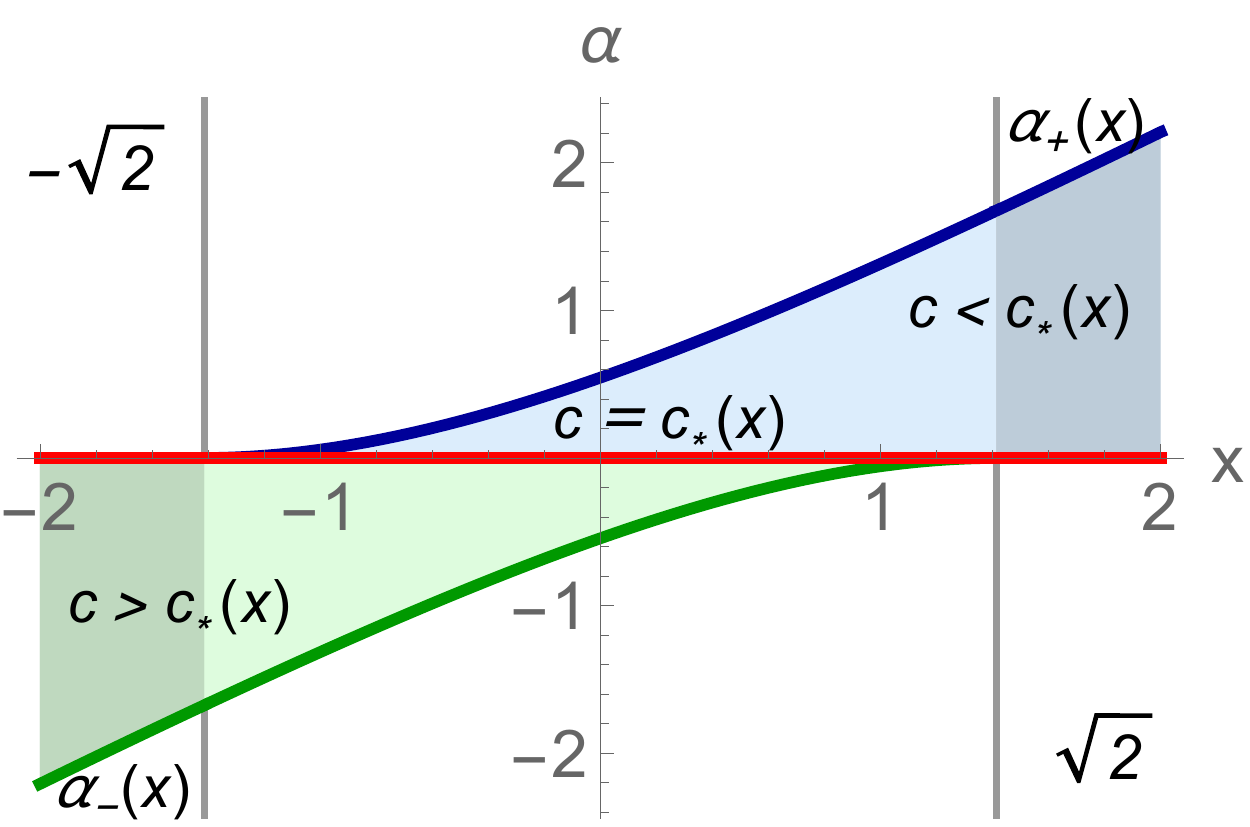}
\caption{Region of physical solutions in the $(x,\alpha)$-plane  enclosed by the lines $\alpha=\alpha_{+}(x)$ (solid blue line) and $\alpha=\alpha_{-}(x)$ (solid green line). As noted in the text, $\alpha=0$ (solid red line) corresponds to the Wigner law and, thus, along this line the fraction of eigenvalues to the left of $x$ must be $c=c_{\star}(x)$. From the ($\alpha=0$)-line and  increasing $\alpha$ we enter into the blue filled region, which is characterised by a constrained spectral density with  $c$ below to the typical value $c_{\star}$. Upon further increasing $\alpha$ we eventually arrive to the condition $\alpha=\alpha_{+}(x)$ corresponding precisely to $c=0$ or, in order words, a constrained spectral density with all the eigenvalues to the right of $x$. Alternatively, starting from $\alpha=0$ and  decreasing $\alpha$ we enter in the green filled region which corresponds to $c>c_{\star}(x)$, until we eventually we arrive to the boundary line $\alpha=\alpha_{-}(x)$. The latter corresponds to $c=1$, that is, a constrained spectral density with all eigenvalues to the left of $x$.}
\label{alphavsx}
\end{figure}

\begin{figure}[h]
\centering
\includegraphics[width=10cm,height=7cm]{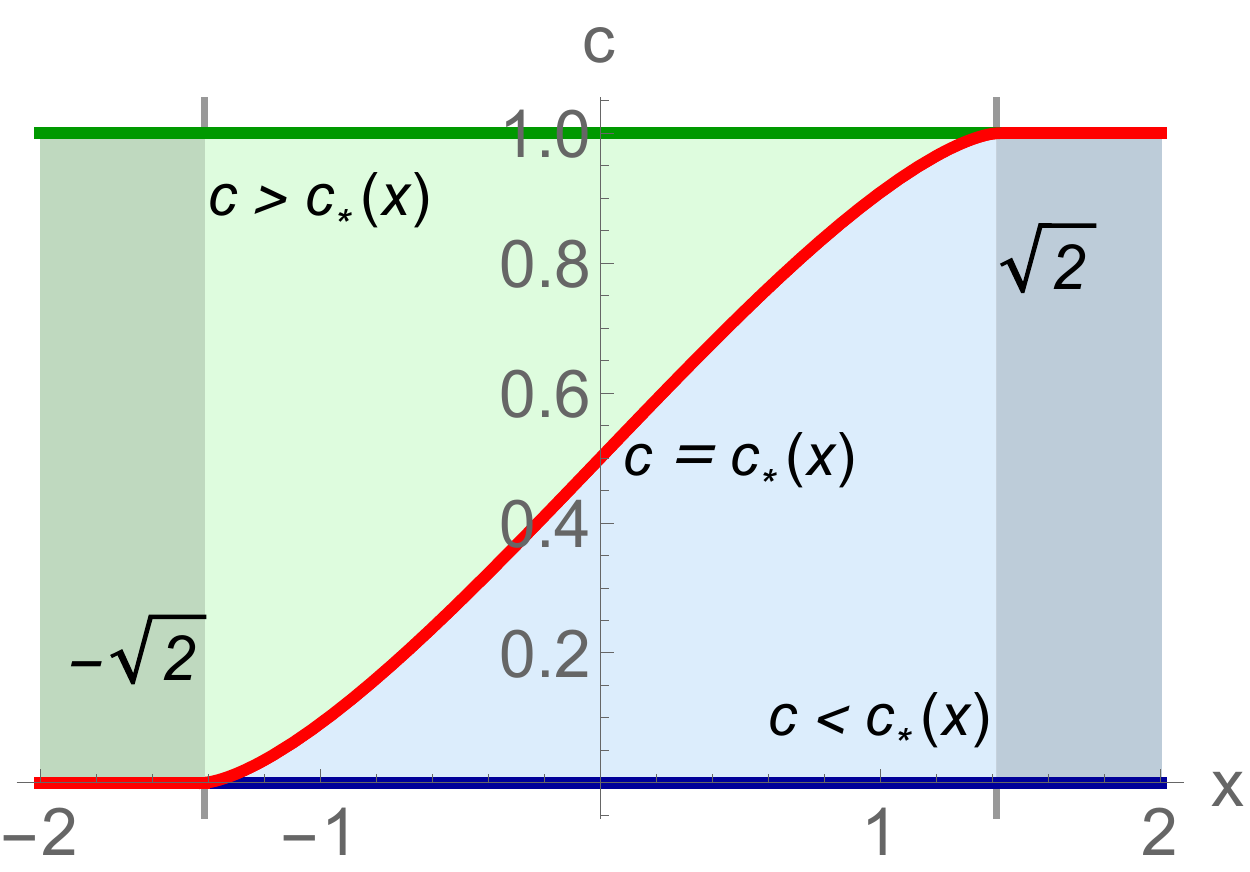}
\caption{The region shown in this plot corresponds to transforming the region of Fig. \ref{alphavsx} into the $(x,c)$-plane. The red solid line corresponds to the function $c_\star(x)$.}
\label{cvsx}
\end{figure}

\begin{figure*}
\includegraphics[width=4.5cm,height=4.5cm]{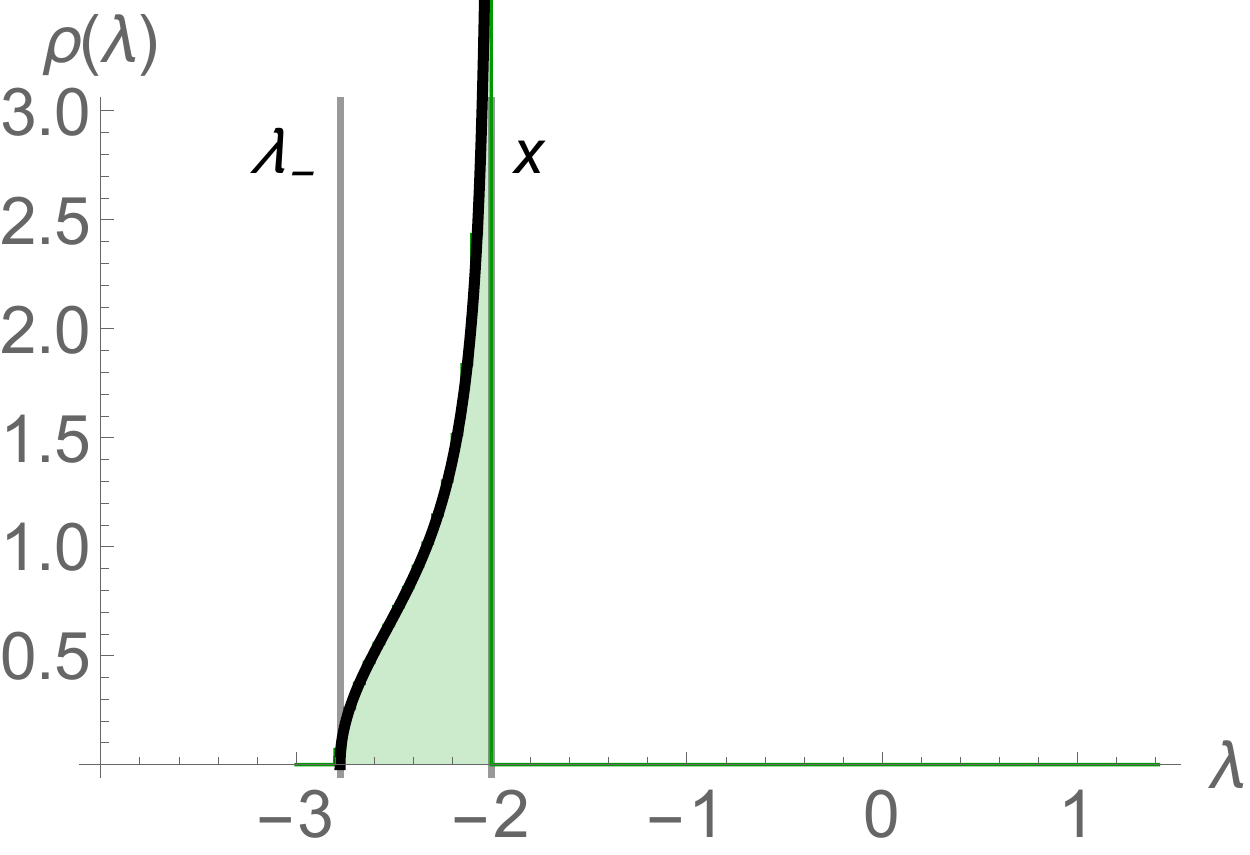}\quad\quad
\includegraphics[width=4.5cm,height=4.5cm]{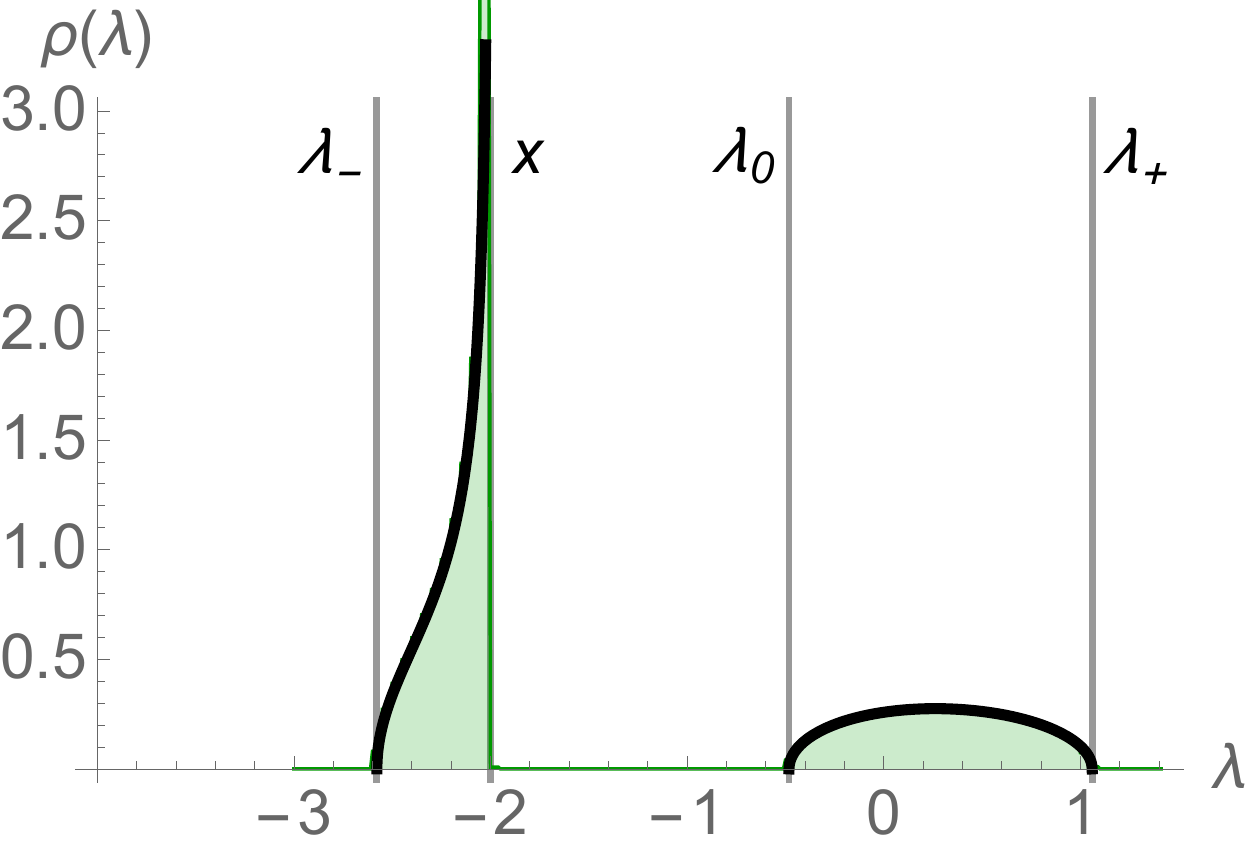}\quad\quad
\includegraphics[width=4.5cm,height=4.5cm]{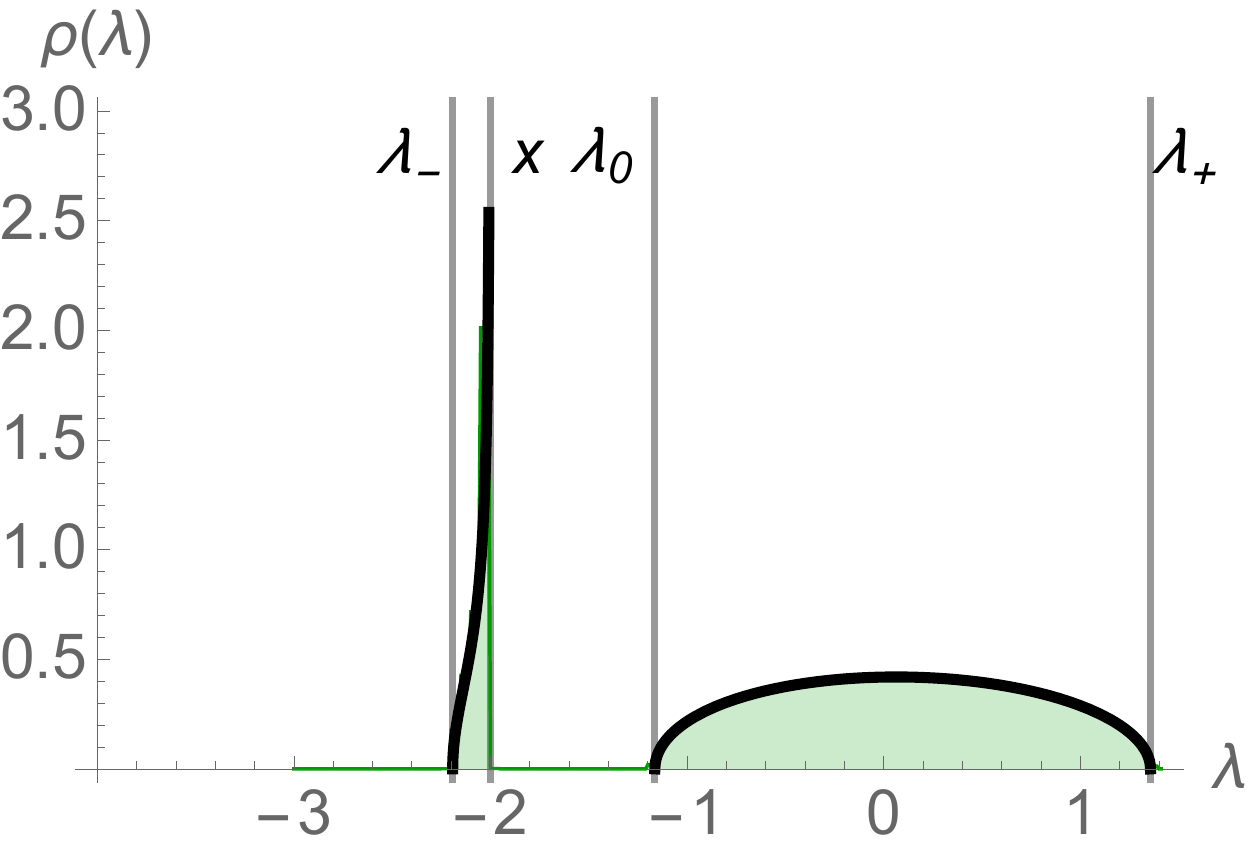}
\caption{Plot of the analytical expressions of the spectral density $\rho(\lambda)$ (solid black line) for $x=-2$ and various values of $c=1,2/3,1/6$. We have compared these formulas with estimates from Monte Carlo simulations of the Coulomb fluid (green filled region below the black lines).}
\label{rho}
\end{figure*}

To solve the saddle-point equations \eref{eq:spe1}, \eref{eq:spe2} and \eref{eq:spe3}   we introduce the Hilbert-Stieltjes transform of the density $\rho(\lambda)$, $S(z)=\int d\lambda \frac{\rho(\lambda)}{z-\lambda}$, with $z\in\mathbb{C}$. Here  $S(z)$  is usually  called the resolvent. This helps to write the derivative of the  the saddle-point equation \eref{eq:spe1} as a  quadratic equation for $S(z)$. Indeed, differentiating with respect to $\lambda$ in   eq.  \eref{eq:spe1} we obtain the  Tricomi equation:
\begin{equation}
\lambda+B_1\delta(x-\lambda)=\dashint d\lambda'\frac{\rho(\lambda')}{\lambda-\lambda'}\,,
\label{app:tricomi}
\end{equation}
with $B_1=- A_1/2$. We then multiply eq. \eref{app:tricomi} by $\rho(\lambda)/(z-\lambda)$ and integrate over $\lambda$, viz.
\begin{equation}
\hspace{-1cm}\int d\lambda\frac{\rho(\lambda)}{z-\lambda}\lambda+B_1\int d\lambda\frac{\rho(\lambda)}{z-\lambda}\delta(x-\lambda)=\int d\lambda\dashint d\lambda'\frac{\rho(\lambda )\rho(\lambda')}{(\lambda-\lambda')(z-\lambda)}\,.
\label{app:t2}
\end{equation}
Next, we must work out the three terms in \eref{app:t2} to write them in terms of $S(z)$. For the first term we write
\begin{equation*}
\int d\lambda\frac{\rho(\lambda)}{z-\lambda}\lambda=\int d\lambda\frac{\rho(\lambda)}{z-\lambda}(\lambda-z)+z\int d\lambda\frac{\rho(\lambda)}{z-\lambda}\lambda=-1+zS(z)\,.
\end{equation*}
For the second term we have:
\begin{equation*}
B_1\int d\lambda\frac{\rho(\lambda)}{z-\lambda}\delta(x-\lambda)=B_1\frac{\rho(x)}{z-x}\,.
\end{equation*}
Finally, for the third term we arrive at
\begin{eqnarray*}
T_3&=&\int d\lambda\dashint d\lambda'\frac{\rho(\lambda )\rho(\lambda')}{(\lambda-\lambda')(z-\lambda)}\\
&&=\int d\lambda\dashint d\lambda'\rho(\lambda )\rho(\lambda')\left[\frac{1}{z-\lambda}+\frac{1}{\lambda-\lambda'}\right]\frac{1}{z-\lambda'}\\
&&=S^2(z)-T_3\,,
\end{eqnarray*}
which implies that $T_3=\frac{1}{2}S^{2}(z)$. After denoting $\alpha=-B_1\rho(x)$ we obtain the following  quadratic equation for $S(z)$
\begin{equation}
\frac{1}{2}S^{2}(z)-zS(z)+1+\frac{\alpha}{z-x}=0\,,
\label{eq:resolvent}
\end{equation}
whose solution is
\begin{equation*}
S_{\pm}(z)=z\pm\sqrt{\frac{P_3(z)}{z-x}}\,,\quad\quad P_3(z)=z^3-xz^2-2z+2(x-\alpha)\,.
\end{equation*}
Here $\alpha$ is a constant to be  determined by the saddle-point equations \eref{eq:spe2} and \eref{eq:spe3} as a function of the paramerters of the problem $c$ and $x$, that is $\alpha=\alpha(c,x)$. In anticipation of the detailed mathematical analysis below we notice that for $\alpha=0$  we  recover the semicircle law. Hence $\alpha(c_{\star}(x),x)=0$ and, as we will see below, whenever $\alpha$ is either negative or positive we are either in the regime $c>c_{\star}(x)$ or in the regime $c<c_{\star}(x)$, respectively.\\
To extract the density of eigenvalues $\rho(\lambda)$ from $S(z)$, we recall that the resolvent $S(z)$ has the following properties: (i) it is analytic everywhere on $\mathbb{C}$ but not on the cuts on the real line where the density is defined; (ii) it must behave as $\frac{1}{z}$ as $|z|\to\infty$; (iii) it is real for real $z$ outside of the domain of the density; (iv)  using distribution theory we know that if we approach a point $\lambda$ belonging to the domain of $\rho(\lambda)$, that is  $S(\lambda\pm i\epsilon)= g(\lambda)\mp i\pi \rho(\lambda)$,  which implies $\rho(\lambda)=-\lim_{\epsilon\to0^{+}}\frac{1}{\pi}{\rm Im}[ S(\lambda+i\epsilon)]$. Property (ii) implies that we must take the  solution  $S_{-}(z)$.\\ 
To elucidate the spectral density $\rho(\lambda)$ one first needs to understand the behaviour of the three roots of $P_3(z)$. These are given by the formulas:
\begin{eqnarray*}
\lambda_{+}&=&\frac{1}{3} \Bigg(\Delta(\alpha,x)+\frac{x^2+6}{\Delta(\alpha,x)}+x\Bigg)\,,\\
\lambda_0&=&\frac{1}{3} \bigg(-\frac{1+i \sqrt{3}}{2} \Delta(\alpha,x)-\frac{1-i\sqrt{3} }{2}\frac{x^2+6}{\Delta(\alpha,x)}+ x\Bigg)\,,\\
\lambda_{-}&=&\frac{1}{3} \Bigg(-\frac{1-i\sqrt{3}}{2} \Delta(\alpha,x)-\frac{1+i \sqrt{3}}{2}\frac{ x^2+6}{\Delta(\alpha,x)}+ x\Bigg)\,,
\end{eqnarray*}
where we have defined
\begin{eqnarray*}
\Delta(\alpha,x)&=&\Big(27\sqrt{(\alpha-\alpha_-(x))(\alpha-\alpha_{+}(x))}+x^3-18 x+27\alpha\Big)^{1/3}\,,\\
\alpha_{\pm}(x)&=&\frac{1}{27}\left(x(18-x^2)\pm (6+x^2)^{3/2}\right)\,.
\end{eqnarray*}
Given a value of $x$, whenever $\alpha\in[\alpha_{-}(x),\alpha_{+}(x)]$ the three roots are real. However, not all possible points in this subregion of the $(x,\alpha)$-plane are physical.  This can be understood by looking at plots of the roots as a function of $\alpha$ for fixed values of $x$, like the ones presented in Fig. \ref{roots}.
Here we can see (as well as algebraically from  the form of the polynomial $P_3(z)$) that when $\alpha=0$, we recover the semicircle law. Then we have two cases: either the position of the barrier is within the natural support of Wigner's law, that is $x\in[-\sqrt{2},\sqrt{2}]$ (corresponding to the middle plot in Fig. \ref{roots}), or  $x\not\in[-\sqrt{2},\sqrt{2}]$ (corresponding to the left and right plots in Fig. \ref{roots}). In the first case ($x\in[-\sqrt{2},\sqrt{2}]$) we have that $\alpha\in[\alpha_{-},\alpha_{+}]$. The value of $\alpha$ controls the fraction of eigenvalues to the left of $x$: when varying $\alpha$, denoted as a black vertical line in Fig. \ref{roots}, we can go from all eigenvalues  within the interval $[\lambda_{-},x]$ (left-most position  corresponding to $c=1$), to all eigenvalues within the interval $[x,\lambda_{+}]$ (right-most position corresponding to $c=0$). At any other position of $\alpha$, the eigenvalues are found distributed between two blobs: either $[\lambda_{-},x]\cup[\lambda_0,\lambda_{+}]$ (corresponding to $\alpha\in[\alpha_{-},0]$) or $[\lambda_{-},\lambda_0]\cup[x,\lambda_{+}]$ (corresponding to $\alpha\in[0,\alpha_{+}]$).\\
On the other side, when $x\not\in[-\sqrt{2},\sqrt{2}]$ (left and right plots in Fig. \ref{roots}) then either $\alpha\in[\alpha_{-},0]$ (corresponding to $x<-\sqrt{2}$, left figure) or $\alpha\in[0,\alpha_{+}]$ (corresponding to $x>\sqrt{2}$, right figure). Consider the first case (the second one works similarly). When $\alpha=\alpha_{-}$, all eigenvalues belong to the interval $[\lambda_{-},x]$, which corresponds to $c=1$. As $\alpha$ moves to the right, the eigenvalues can be found within the two blobs $[\lambda_{0},x]\cup[\lambda_0,\lambda_{+}]$ until, finally for $\alpha=0$ we recover Wigner's law. The latter corresponds to having a zero fraction of eigenvalues $c=0$ to the left of $x$.\\
These restrictions are summarised in Fig. \ref{alphavsx}. Here we have redefined the solutions of $\alpha_{\pm}(x)$ coming from the mathematical analysis but adding in the previous the physical interpretation:
\begin{equation*}
\alpha_{-}(x)=\left\{
\begin{array}{ll}
\frac{1}{27}\left(x(18-x^2)- (6+x^2)^{3/2}\right)& x\leq\sqrt{2}\\
0&x>\sqrt{2}
\end{array}
\right.\,,
\end{equation*}
and
\begin{equation*}
\alpha_{+}(x)=\left\{
\begin{array}{ll}
0&x<-\sqrt{2}\\
\frac{1}{27}\left(x(18-x^2)+ (6+x^2)^{3/2}\right)& x\geq-\sqrt{2}
\end{array}
\right.\,.
\end{equation*}
In  \fref{alphavsx} we have plotted the region of physical solutions in the $(x,\alpha)$-plane, which is  encapsulated by the lines $\alpha=\alpha_{+}(x)$ (solid blue line  in Fig. \ref{alphavsx}) and $\alpha=\alpha_{-}(x)$ (solid green line  in Fig. \ref{alphavsx}).  Recall that if we  follow  along the line $\alpha=0$ (solid red line in Fig. \ref{alphavsx})  we have that the fraction of eigenvalues to the left of $x$ should be $c=c_{\star}(x)$. From $\alpha=0$ and  increasing $\alpha$, we are exploring those constrained spectral densities with $c<c_{\star}(x)$ (corresponding to the blue filled region in Fig. \ref{alphavsx}). Upon further increasing $\alpha$ we will eventually arrive to the line $\alpha=\alpha_{+}(x)$, which  corresponds precisely to $c=0$ or, in order words, a constrained spectral density with all the eigenvalues to the right of $x$. Alternatively, starting from $\alpha=0$ and decreasing the value of $\alpha$ we explore constrained spectral densities with  $c>c_{\star}(x)$ (green filled region in Fig. \ref{alphavsx}) until we eventually we arrive to the boundary line $\alpha=\alpha_{-}(x)$. The latter corresponds to $c=1$ or to a constrained spectral density with all eigenvalues to the left of $x$.\\
This analysis relies  on the  parameter $\alpha$, which lacks a direct physical interpretation. To switch the analysis  into the $(x,c)$-plane we first need to derive $c_\star(x)=\int_{-\sqrt{2}}^{x} dy\rho_{{\rm sc}}(y)$. We arrive at
\begin{equation}
c_{\star}(x)=\left\{
\begin{array}{ll}
0&x<-\sqrt{2}\\
\frac{\pi+x\sqrt{2-x^2}+2{\rm arcsin}\left(\frac{x}{\sqrt{2}}\right)}{2\pi}& x\in[-\sqrt{2},\sqrt{2}]\\
1&x>\sqrt{2}
\end{array}
\right.\,.
\label{eq:ctyp}
\end{equation}
The region of physical solutions in the $(x,\alpha)$-plane (in Fig. \ref{alphavsx}) is transformed into that reported in Fig. \ref{cvsx} for the $(x,c)$-plane.\\
Taking into account all the nuances we can finally write a constrained spectral density for both regimes. For $c>c_{\star}(x)$ we have that
\begin{equation*}
\hspace{-2.5cm}\rho(\lambda)=\frac{1}{\pi}\sqrt{\frac{(\lambda_{+}-\lambda)(\lambda_{0}-\lambda)(\lambda-\lambda_{-})}{x-\lambda}}\openone_{\lambda\in[\lambda_{-},x]}+\frac{1}{\pi}\sqrt{\frac{(\lambda_{+}-\lambda)(\lambda-\lambda_{0})(\lambda-\lambda_{-})}{\lambda-x}}\openone_{\lambda\in[\lambda_{0},\lambda_{+}]}\,,
\end{equation*}
while for $c<c_\star(x)$ we write instead
\begin{equation*}
\hspace{-2.5cm}\rho(\lambda)=\frac{1}{\pi}\sqrt{\frac{(\lambda_{+}-\lambda)(\lambda_{0}-\lambda)(\lambda-\lambda_{-})}{x-\lambda}}\openone_{\lambda\in[\lambda_{-},\lambda_0]}+\frac{1}{\pi}\sqrt{\frac{(\lambda_{+}-\lambda)(\lambda-\lambda_{0})(\lambda-\lambda_{-})}{\lambda-x}}\openone_{\lambda\in[x,\lambda_{+}]}\,.
\end{equation*}
Here $\openone_{x\in[a,b]}$ is an indicator function equal to the unity if $x\in[a,b]$ or zero otherwise. In particular, we recover the well-known expressions of the spectral density for all eigenvalues to the left of the barrier ($c=1$) and all eigenvalues to the right of the barrier ($c=0$),  which we denote as $\rho_{{\rm L}}(\lambda)$  and $\rho_{{\rm R}}(\lambda)$, respectively. This would correspond to following the lines $\alpha_{-}(x)$ and $\alpha_{+}(x)$ in the $(x,\alpha)$-plane, respectively. Their formulas are
\begin{equation*}
\rho_{{\rm L}}(\lambda)=\left\{
\begin{array}{ll}
\frac{1}{\pi}\sqrt{\frac{\lambda-a_{-}(x)}{x-\lambda}}\left|\lambda-b_{+}(x)\right|\openone_{\lambda\in[a_{-}(x),x]}&x\leq \sqrt{2}\\
\rho_{sc}(\lambda)& x>\sqrt{2}
\end{array}\right.\,,
\end{equation*}
and
\begin{equation*}
\rho_{{\rm R}}(\lambda)=\left\{
\begin{array}{ll}
\rho_{sc}(\lambda)& x<-\sqrt{2}\\
\frac{1}{\pi}\sqrt{\frac{\lambda-a_{+}(x)}{x-\lambda}}\left|\lambda-b_{-}(x)\right|\openone_{\lambda\in[x,a_{+}(x)]}& x\geq-\sqrt{2}
\end{array}
\right.\,,
\end{equation*}
with $a_{\pm}(x)=\frac{1}{3}(x\pm2\sqrt{6+x^2})$ and $b_{\pm}(x)=\frac{1}{3}(x\pm\sqrt{6+x^2})$.\\
To get rid of the parameter $\alpha$ completely we  must derive a function  $\alpha(c,x)$.
\begin{figure}[h]
\centering
\includegraphics[width=10cm,height=7cm]{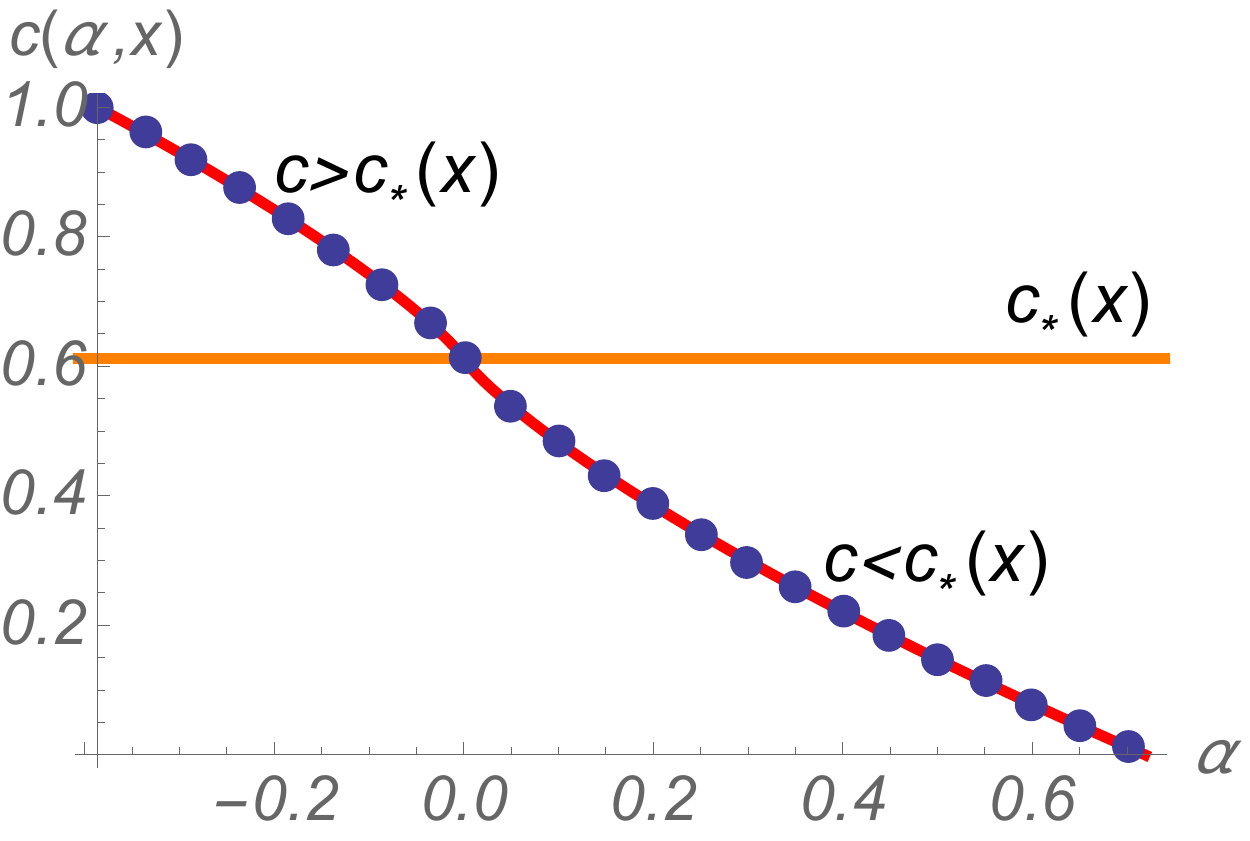}
\caption{Plot of the analytical formulas for $c(\alpha,x)$ (solid red line) for $x=1/4$ and comparison with numerics (solid blue circles). The constant solid orange line $c_{\star}(x)$ corresponds to the typical value given according to eq. \eref{eq:ctyp} for $x=1/4$.}
\label{functionc}
\end{figure}
 We are only able to provide this function implicitly by working out  an expression for $c(\alpha,x)$  from the condition $c=\int_{-\infty}^{x}d\lambda\,\rho(\lambda)$. Using complete elliptic integrals we find that for $c>c_{\star}(x)$, the function $c(\alpha,x)$ takes the following form
\begin{eqnarray*}
c(\alpha,x)&&=\frac{1}{2\pi \sqrt{(a-c) (b-d)}}\Bigg[4 (a-d)  \Pi \left(\frac{d-c}{a-c}\right)\\
&&+(a-c)\left\{ c(b-d) E-(a-d)(a-b+c) K\right\}\Bigg]\,,
\end{eqnarray*} 
with $a=\lambda_{+}$, $b=\lambda_{0}$, $c=x$, and $d=\lambda_{-}$. For $c<c_{\star}(x)$, the function $c(\alpha,x)$ reads
\begin{eqnarray*}
c(\alpha,x)&&=\frac{1}{2\pi \sqrt{(a-c) (b-d)}}\Bigg[4 (b-c) \Pi \left(\frac{c-d}{b-d}\right)\\
&&+ (b-d)\left\{b(a-c)  E-(b-c) (a-2c-2d) K\right\}\Bigg]\,,
\end{eqnarray*} 
where $a=\lambda_{+}$, $b=x$, $c=\lambda_{0}$, and $d=\lambda_{-}$  (see \ref{app:ca}). Here $K\equiv K(k)$,  $E\equiv E(k)$, and $\Pi(n)\equiv\Pi(n,k)$  are the complete elliptic integrals of the first, second, and third kind, respectively, with elliptic modulus $k=\sqrt{\frac{(a-b) (c-d)}{(a-c) (b-d)}}$. In Fig. \ref{functionc} we have plotted the function $c(\alpha,x)$ versus $\alpha$ for a fixed value of $x$. We have also compared the analytical results with the numerical evaluation of $c=\int_{-\infty}^{x}d\lambda\,\rho(\lambda)$.\\
Once we have $\alpha(c,x)$, we can plot the constrained spectral density $\rho(\lambda)$ for varying values of $c$ by fixing a value of $x$. This is done in Fig. \ref{rho} where we show   $\rho(\lambda)$ in the region $c>c_{\star}(x)$ (corresponding to the green filled region in Fig. \ref{cvsx}). In these plots,  we have fixed the value of $x=-2$ and plot $\rho(\lambda)$ along this vertical line in Fig. \ref{cvsx} for three values of $c=1,2/3,1/6$. Notice that the first value $c=1$ corresponds precisely to being on the top of the horizontal dark green line in Fig. \ref{cvsx} (or, equivalenty, on top of $\alpha=\alpha_{+}(x)$ of Fig. \ref{alphavsx}), which implies that that constrained spectral density is precisely given by $\rho_{{\rm L}}(\lambda)$.
\subsection{Determining  expressions for the second moment of the spectral density and constants $A_1$ and $A_2$}
\begin{figure*}
\includegraphics[width=4.5cm,height=4.5cm]{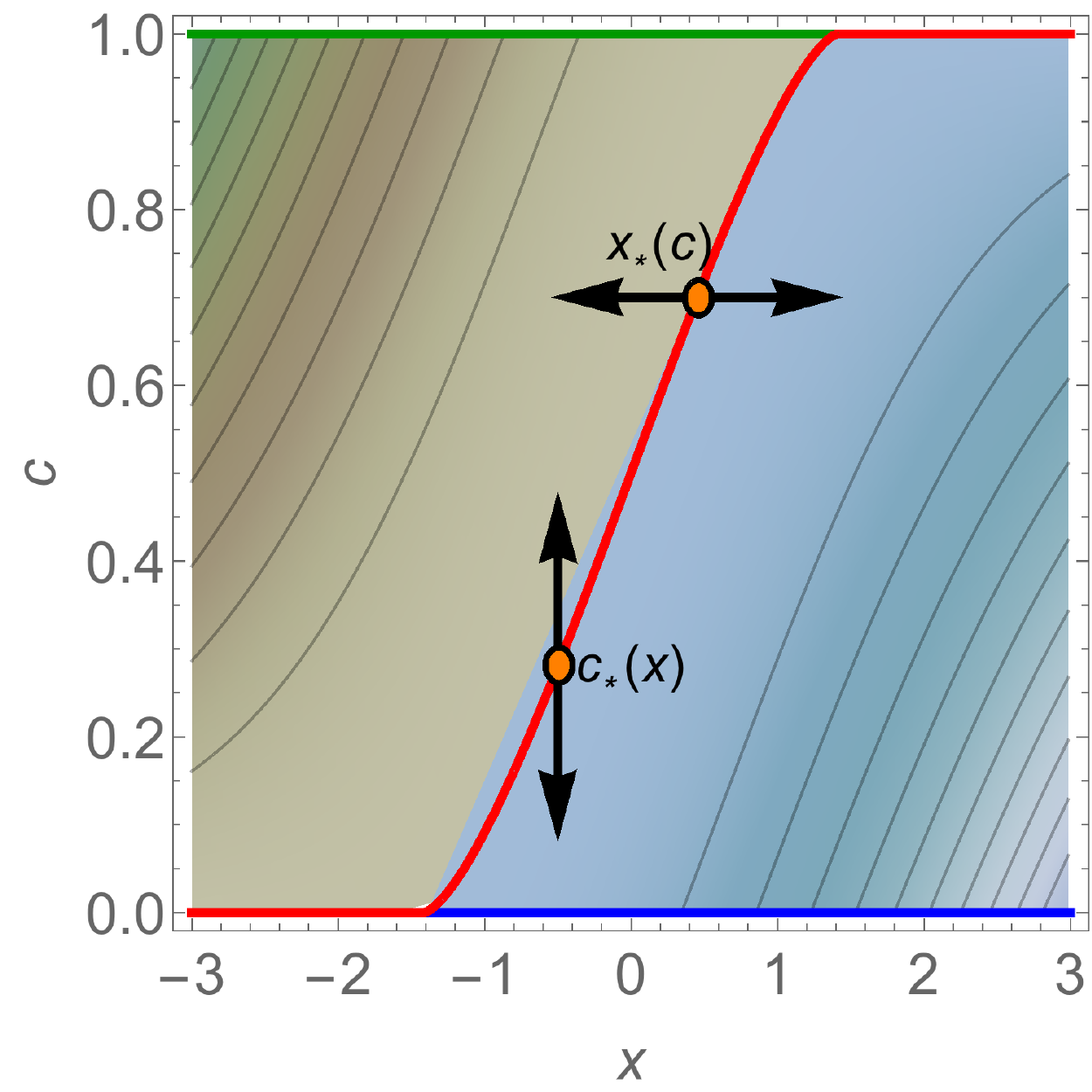}\quad\quad
\includegraphics[width=4.5cm,height=4.5cm]{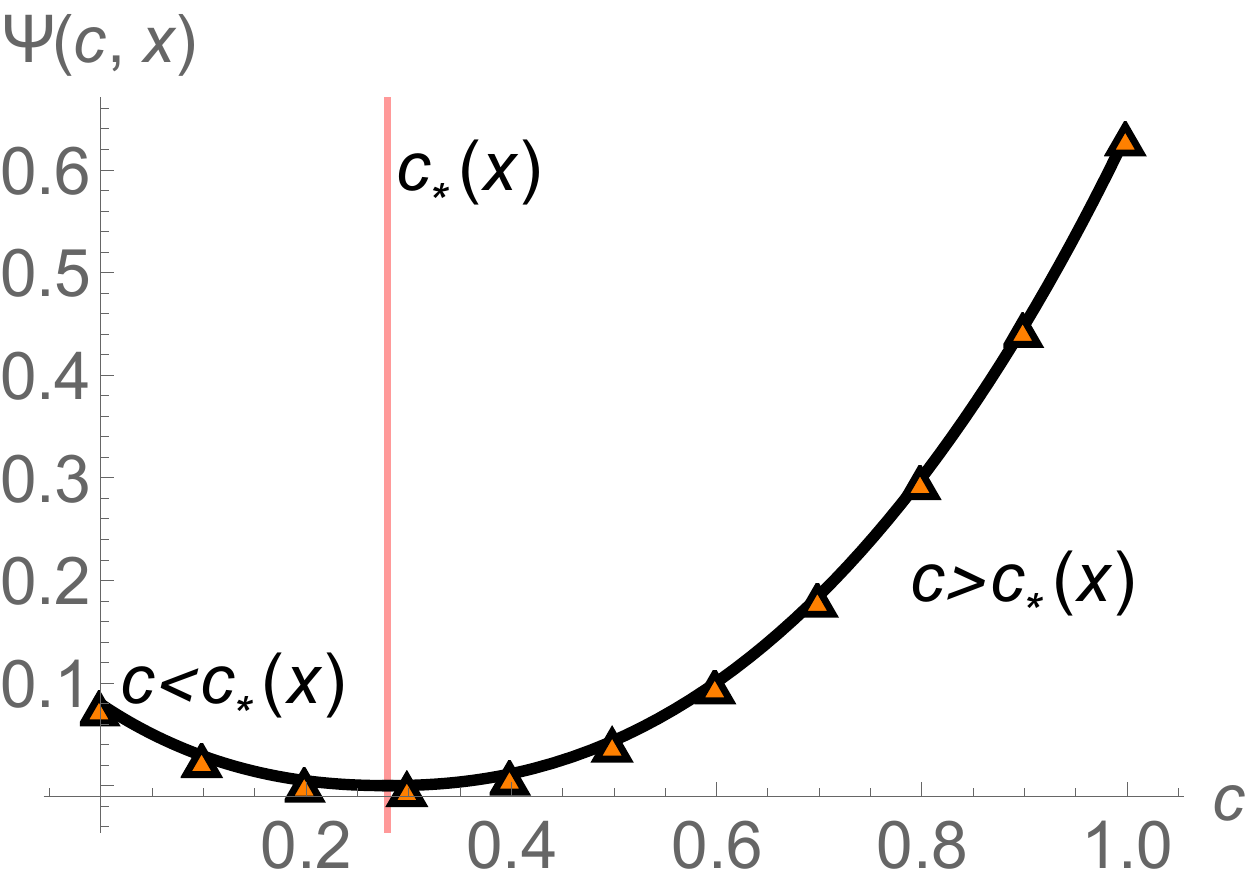}\quad\quad\quad
\includegraphics[width=4.5cm,height=4.5cm]{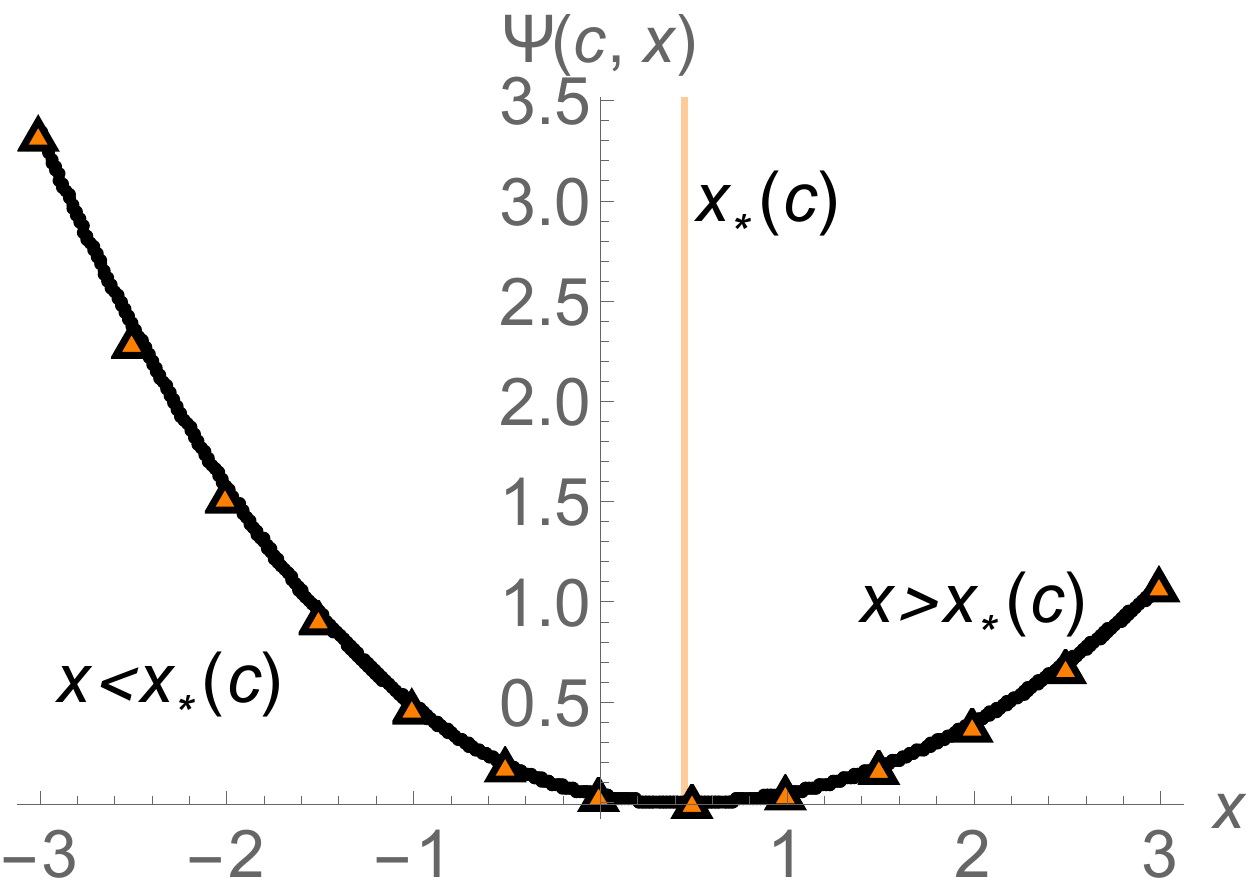}
\caption{Left: density plot of $\Psi(c,x)$. We also plot the line $c_{\star}(x)$ along which the rate function is zero. Middle: plot of the rate function $\Psi(c,x)$ against $c$ for a given value of $x=-1/2$. We have also compared the theoretical results (solid black line) with Monte Carlo simulations of the Coulomb fluid taking $N=600$ (triangles). Right: plot of the rate function $\Psi(c,x)$ against $x$ for a given value of $c=0.7$. We have compared with Monte Carlo simulations with $N=500$ (triangles). }
\label{rate}
\end{figure*}

We move on to derive formulas for the three terms appearing in \eref{eq:action2}, which will allow us to arrive at a final expression for the rate function \eref{eq:rf}. We start by recalling that the resolvent can be expanded as $S_{-}(z)=\sum_{n=0}^\infty\frac{\mu_n}{z^{n+1}}$  with  $\mu_n=\int\rho(\lambda)\lambda^n$. Using our expression for the resolvent $S_{-}(z)$ automatically yields
\begin{equation*}
\int d\lambda \rho(\lambda)\lambda^2=\frac{1}{2}+\alpha x\,.
\end{equation*}
To find expressions for  the constants $A_1$ and $A_2$ we need to treat the cases $c>c_{\star}(x)$ and $c<c_{\star}(x)$ separately.\\
For $c>c_{\star}(x)$ the domain of the density is $[\lambda_{-},x]\cup[\lambda_0,\lambda_{+}]$.  We then  evaluate the  saddle-point equation \eref{eq:spe1}  at the following three points inside this domain: $\lambda_{+}-\epsilon$, $\lambda_0+\epsilon$ and $x-\epsilon$ for $\epsilon\to 0^{+}$. They yield the following 3 equations
\begin{eqnarray*}
\lambda^2_{+}+A_2&=&2\int d\lambda' \rho(\lambda')\log|\lambda_{+}-\lambda'|\,,\\
\lambda^2_{0}+A_2&=&2\int d\lambda' \rho(\lambda')\log|\lambda_{0}-\lambda'|\,,\\
x^2+A_1+A_2&=&2\int d\lambda' \rho(\lambda')\log|x-\lambda'|\,,
\end{eqnarray*}
which can be combined to arrive at  two formulas for $A_1$ and $A_2$:
\begin{eqnarray}
A_1&=&\lambda^2_{0}-x^2+2\int d\lambda' \rho(\lambda')\log\left|\frac{x-\lambda'}{\lambda_{0}-\lambda'}\right|\,,\\
A_2&=&-\lambda^2_{+}+2\int d\lambda' \rho(\lambda')\log|\lambda_{+}-\lambda'|\,.
\label{app:constants}
\end{eqnarray}
This, in turn, can be written in terms of the resolvent as
\begin{eqnarray*}
A_1&=&\lambda_0^2-x^2-2\int^{\lambda_0}_{x} dz S_{+}(z)\,,\\
A_{2}&=&-\lambda_{+}^2+2\left[\log(\lambda_{+})-\int_{\lambda_{+}}^\infty dz\left(S_{-}(z)-\frac{1}{z}\right)\right]\,.
\end{eqnarray*}
For $c<c_{\star}(x)$, the support of the spectral density is  $[\lambda_{-},\lambda_0]\cup[x,\lambda_{+}]$. Again, we evaluate the saddle-point equation \eref{eq:spe1} at the points $\lambda_{+}-\epsilon,x+\epsilon,\lambda_0-\epsilon$, for $\epsilon\to 0^+$:
\begin{eqnarray*}
&&2\int d\lambda' \rho(\lambda')\log|\lambda_{+}-\lambda'|= \lambda^2_++A_2\,,\\
&&2\int d\lambda' \rho(\lambda')\log|x-\lambda'|= x^2+A_2\,,\\
&&2\int d\lambda' \rho(\lambda')\log|\lambda_{0}-\lambda'|= \lambda^2_{0}+A_1+A_2\,,
\end{eqnarray*}
which, after combining and expressing in terms of the resolvent, yield
\begin{eqnarray*}
A_1&=&x^2-\lambda_0^2-2\int_{\lambda_0}^{x}dz S_{-}(z)\,,\\
A_{2}&=&-\lambda_{+}^2+2\left[\log(\lambda_{+})-\int_{\lambda_{+}}^\infty dz\left(S_{-}(z)-\frac{1}{z}\right)\right]\,.
\end{eqnarray*}
It is not hard to see that the formulas we have found for $A_1$ and $A_2$ in both regimes  $c>c_{\star}(x)$ and $c<c_{\star}(x)$ are actually the same. Finally, we can write a compact expression for the rate function $\Psi(c,x)$:
\begin{eqnarray}
\hspace{-2cm}\Psi(c,x)&&=\frac{1}{2}\Bigg[\frac{1}{2}\left(\frac{1}{2}+\alpha x\right)+\frac{\lambda_{+}^2}{2}-\log(\lambda_{+})+\int_{\lambda_{+}}^\infty dz\left(S_{-}(z)-\frac{1}{z}\right)-\frac{3+2\log(2)}{4}\nonumber\\
\hspace{-2cm}&&+\frac{c}{2}\left(2\int_{\lambda_0}^{x} dz S_{-}(z)+\lambda_0^2-x^2\right)\Bigg]\,.
\label{ratere}
\end{eqnarray}
As we  will discuss later, the expression \eref{ratere} is useful to analyse deviations around the typical line $(x,c_\star(x))$ in the $(x,c)$-plane. Still, for general values of $(x,c)$ we need to evaluate the integrals appearing in \eref{ratere}. After a lengthy derivation one eventually reaches (see \ref{app:ei}) the following formula:
\begin{equation*}
\Psi(c,x)=\frac{1}{2}\left[\frac{1}{2}\left(\frac{1}{2}+\alpha x\right)-\frac{A_1}{2}c-\frac{A_2}{2}-\Omega_0\right]\,,
\end{equation*}
where the constants $A_1$ and $A_2$ can be written in terms of incomplete elliptic integrals. Indeed, for $c>c_{\star}(x)$  we have 
\begin{eqnarray}
A_1&=&-2I_1(\lambda_+,\lambda_0,x,\lambda_{-})\,,\\
A_2&=&-\lambda^2_{+}+2\left(\log\lambda_++I_2(\lambda_{+},\lambda_0,x,\lambda_-)\right)\,,\\
I_1&=&\frac{1}{2 \sqrt{(a-c) (b-d)}}\Bigg[4 (a-b) \Pi' \left(\frac{b-c}{a-c}\right)\nonumber\\
&&+ (a-c) \left[(b-a) (a+c-d) K'+c (d-b) E'\right]\Bigg]\,,\\
I_2&=&\frac{1}{2 \sqrt{(a-c) (b-d)}}\Big[4 (a-b)\Pi' \left(\theta,\frac{b-c}{a-c}\right)\nonumber\\
&+&(a-c)[(b-a)  (a+c-d) F'(\theta )+c(d-b) E'(\theta)]\Big]\nonumber\\
&+&\frac{1}{2} \left(d(b-c) +1-2 \log \left(\frac{2 a}{-d}\right)\right)\,,
\end{eqnarray}
with definitions  $a=\lambda_{+}$, $b=\lambda_{0}$,  $c=x$, and  $d=\lambda_{-}$. For $c<c_{\star}(x)$ we find instead 
\begin{eqnarray}
A_{1}&=&2J_1(\lambda_+,x,\lambda_0,\lambda_-)\,,\\
 A_2&=&-\lambda^2_{+}+2\left(\log(\lambda_+)+J_2(\lambda_+,x,\lambda_0,\lambda_-)\right)\,,\\
J_1&=&\frac{1}{2\sqrt{(a-c)(b-d)}}\Bigg[4 (c-d)\Pi' \left(\frac{b-c}{b-d}\right)\nonumber\\
&-& (b-d)[(c-d) (a-b-d) K'-b (a-c) E']\Bigg]\,,\\
J_{2}&=&\frac{1}{2 \sqrt{(a-c) (b-d)}}\Bigg[4 (a-b)\Pi' \left(\theta,\frac{b-c}{a-c}\right) \nonumber\\
&-&(a-c)[(a-b) (a+b-d) F'(\theta)+ b  (b-d) E'(\theta)]\Bigg]\nonumber\\
&+&\frac{1}{2} \left[1- \log \left(4\right)+(b-c) (b-d)\right]\,,
\end{eqnarray}
with definitions $a=\lambda_{+}$, $b=x$, $c=\lambda_{0}$, and $d=\lambda_{-}$. In both cases $F'(\theta)\equiv F(\theta,k')$, $E'(\theta)\equiv E(\theta,k')$, and $\Pi'(\theta,n)\equiv \Pi(\theta,n,k')$ are the incomplete elliptic integrals of the first, second, and third kind, respectively, with $k'=\sqrt{1-k^2}$, elliptic modulus $k=\sqrt{\frac{(a-b) (c-d)}{(a-c) (b-d)}}$, and argument $\theta=\sin^{-1}\sqrt{\frac{(b-d)}{(a-d)}}$.\\
In fig. \ref{rate} we show plots of the rate function $\Psi(c,x)$. We note that a few points are in order.
Firstly the rate function is zero along the line $c_{\star}(x)$, that is $\Psi(c_{\star}(x),x)=0$, as expected, as this case corresponds to the unperturbed Coulomb fluid configuration of Wigner's semicircle law. Any perturbation of this configuration will increase the energy of the Coulomb fluid, increasing the value of the rate function. This is shown in Fig. \ref{rate}, where we present two cuts of the rate function. In the middle plot of Fig. \ref{rate} we have fixed the position $x$ of the barrier and varied the fraction $c$. Here the rate function is zero for $c_{\star}(x)$, that is, the value of $c$ corresponding to Wigner's law. In this setup, and if we want to keep the position of the barrier constant, there are only two ways to perturbed Wigner's law: either we pull eigenvalues to the left of the barrier ($c>c_{\star}(x)$) or we pull eigenvalues to the right of the barrier ($c<c_{\star}(x)$). Similarly, in the right plot in Fig. \ref{rate} we have instead fixed the fraction $c$ and varied the position of the barrier. Here the rate function is zero at $x_{\star}(c)$, that is, the position of the barrier that will give precisely the fraction $c$ in the semicircle law (i.e. $x_{\star}(c)$ is the inverse function of $c_{\star}(x)$). While keeping $c$ constant, there are two ways to perturbed Wigner's law: either by pushing the barrier to the left $(x<x_{\star}(c))$ or to the right $(x>x_{\star}(c))$.
\section{Tail Cumulative Distribution Function of $c$}
\label{sec:tcd}
The tail CDF of $\varrho(c)=e^{-\beta N^2\Psi(c,x)}$ reads
\begin{equation*}
\overline{F}_{\mathcal{C}_x}(c)={\rm Prob}[\mathcal{C}_x\geq c]=\int_{c}^1 dc'\varrho(c')\,.
\end{equation*}
Since  $\varrho(c)$ is peaked at $c=c_\star(x)$ then we have that\footnote{If the rate function is one-branched the same reasoning obviously applies.} :
\begin{eqnarray*}
\overline{F}_{\mathcal{C}_x}(c)&=&e^{-\beta N^2\Psi(c,x)}\,,\quad c>c_{\star}(x)\\
F_{\mathcal{C}_x}(c)&=&e^{-\beta N^2\Psi(c,x)}\,,\quad c<c_{\star}(x)\,.
\end{eqnarray*}
Once the CDF of the SIN is found, we automatically arrive at the corresponding CDF for the $k$-th eigenvalue, \textit{viz.}
\begin{eqnarray}
F_{\lambda_k}(x)&=&e^{-\beta N^2\Psi(k/N,x)}\,,\quad x<x_{\star}(k/N)\nonumber\\
\overline{F}_{\lambda_k}(x)&=&e^{-\beta N^2\Psi(k/N,x)}\,,\quad x>x_{\star}(k/N)
\label{eq:LDk}
\end{eqnarray}
Thus, the rate function $\Psi(c,x)$ has a two-fold meaning: as a function of $c$ (for $x$ fixed) it gives information about the large deviations of the SIN; as a function of $x$ (for $c=k/N$ fixed) it gives the large deviations of the $k$-th eigenvalue. Importantly, and as we show below, we find that for extreme eigenvalues, that is for $k=1$ (smallest) or $k=N$ (largest), one is able to derive correctly their left and right large deviation functions within the Coulomb fluid picture. 
\section{Large deviation functions for extreme eigenvalues}
\label{sec:ldf}
Let us focus on the smallest eigenvalues $k=1$\footnote{The largest eigenvalue has identical properties  due to the symmetry of eigenvalues around zero of the Gaussian ensemble.}. This corresponds to study the behaviour of the rate function $\Psi(c,x)$ along the horizontal line $c=0$ (left plot in Fig. \ref{rate}). We have  two distinct cases corresponding to  either $x>-\sqrt{2}$ (right rate function, which we  denote as $\Psi_{+}(x)$) or $x<-\sqrt{2}$ (left rate function, which we  denote as $\Psi_-(x)$). The right rate function is  obtained by taking the limit $\Psi_{+}(x)=\lim_{c\to 0^{+}}\Psi(c,x)$ so that $\overline{F}_{\lambda_1}(x)=e^{-\beta N^2\Psi_{+}(x)}$  for $x>-\sqrt{2}$. In this case, which corresponds to  $0=c<c_{\star}(x)$, we have that $\lambda_{+}>x>\lambda_0=\lambda_{-}$ with $\lambda_{+}=a_+(x)$ and $\lambda_0=\lambda_{-}=b_-(x)$. Looking at the corresponding expressions for $c<c_{\star}(x)$, only the constant $A_2$ contributes in this limit or, equivalently the function $J_{2}(\lambda_{+},x,\lambda_0,\lambda_-)$. Besides, in this limit the elliptic modulus is zero (or the complementary elliptic modulus is one), so that the incomplete elliptic integrals appearing in $J_2$ take the following forms:
\begin{eqnarray*}
F(\theta,1)&=&\log(\tan\theta+\sec\theta)\,,\quad E(\theta,1)=\sin\theta\,,\\
\Pi(\theta,\tilde{\alpha}^2,1)&=&\frac{\log(\tan\theta+\sec\theta)-\tilde{\alpha} \log\sqrt{\frac{1+\tilde{\alpha}\sin\theta}{1-\tilde{\alpha}\sin\theta}}}{1-\tilde{\alpha}^2}\,.
\end{eqnarray*}
Plugging this result back into $J_2$, and after some lengthy algebra, we  obtain:
\begin{equation*}
\hspace{-1.5cm}J_{2}(a,b,c,d)=\frac{1}{18} \Bigg(x \left(\sqrt{x^2+6}-x\right)-18 \log \left(x \left(\sqrt{x^2+6}+x\right)+4\right)+15\Bigg)\,.
\end{equation*}
This helps us to arrive at the following expression for the  $A_{2}$
\begin{eqnarray*}
A_{2}&=&-\frac{1}{9}\left(x+2\sqrt{6+x^2}\right)^2+2\log\left[\frac{1}{3}\left(x+2\sqrt{6+x^2}\right)\right]\\
&&+\frac{1}{9} \Bigg(15+x \left(\sqrt{x^2+6}-x\right)-18\log \left(x \left(\sqrt{x^2+6}+x\right)+4\right)\Bigg)\,,
\end{eqnarray*}
while the action $S_0(c,x)$ takes the form
\begin{eqnarray*}
\hspace{-2cm}S_0(c,x)&=&\frac{3}{4}+\frac{1}{54}\Bigg[x^2(36-x^2)+x(15+x^2)\sqrt{6+x^2} \\
&+&54\log(3)-54\log\left(x+2\sqrt{6+x^2}\right)+54 \log \left(x \left(\sqrt{x^2+6}+x\right)+4\right)\Bigg]\,.
\end{eqnarray*}
After some final standard manipulations of the logarithms we  arrive at
\begin{eqnarray*}
\Psi_{+}(x)&=&\frac{1}{108} \Bigg[-x^4+36 x^2+\sqrt{x^2+6} \left(x^3+15 x\right)\\
&+&27 \left(\log (18)-2 \log \left(\sqrt{x^2+6}-x\right)\right)\Bigg]\,,
\end{eqnarray*}
as reported in \cite{Dean2008}.\\
To derive the left rate function we first realise that $\lim_{c\to 0^{+}}\Psi(c,x)=0$ for $x\leq-\sqrt{2}$. Hence it becomes relevant  to see how the rate function vanishes with $c$ in this limit. Before doing any derivation let us imagine what would happen if $\Psi(c,x)= c\Psi_{-}(x)+\cdots$. Then as $c=1/N$ we  have that $F_{\lambda_1}(x)=e^{-\beta N^2\Psi(k/N,x)}=e^{-\beta N\Psi_{-}(x)}$, which is  the correct scaling with $N$ we expect  for the deviation of the the smallest eigenvalue to the left of its typical value $-\sqrt{2}$.\\
This is precisely what happens mathematically and what allows us to recover the left rate function within the Coulomb fluid picture. To obtain the precise expression for $\Psi_-(x)$ we need to do an expansion around  $c=0^{+}$ for $x<-\sqrt{2}$. It is helpful in this case to realise that this is equivalent of doing the expansion around $\alpha=0$, since we are close to the typical value of $c_{\star}(x)$ for $x<-\sqrt{2}$. There are two ways to tackle the expansion: either directly in the exact expressions or using the expression \eref{ratere} as a starting point. The latter turns out to be easier. To proceed we first need to find expressions of the roots of $P_{3}(x)$ for small $\alpha$. This can be easily achieved by treating the equation $P_3(z)=0$  perturbatively, which yields:
\begin{eqnarray*}
\lambda_{+}&=&\sqrt{2}-\frac{2\alpha}{-4+2\sqrt{2} x}+\cdots\,,\\
\lambda_{0}&=&-\sqrt{2}-\frac{2\alpha}{-4-2\sqrt{2} x}+\cdots\,,\\
\lambda_{-}&=&x-\frac{2\alpha}{2-x^2}+\cdots\,.
\end{eqnarray*}
Using this result  in eq. \eref{ratere}  we arrive at
\begin{eqnarray*}
\Psi(c,x)&=&\frac{1}{2}\left[\frac{1}{2}\left(\frac{1}{2}+\alpha x\right)+\frac{c}{2}\left(2\int_{\lambda_0}^{x} dz S_{-}(z)+\lambda_0^2-x^2\right)+\frac{\lambda_{+}^2}{2}-\log(\lambda_{+})\right.\\
&&+\left.\int_{\lambda_{+}}^\infty dz\left(S_{-}(z)-\frac{1}{z}\right)-\frac{3+2\log(2)}{4}\right]\\
&=&\frac{1}{2}\Bigg[\frac{1}{2}\alpha x+\frac{c}{2}\left(-x \sqrt{x^2-2}-2 \log \left(\sqrt{x^2-2}-x\right)+\log (2)\right)\\
&&+\frac{\alpha}{\sqrt{x^2-2}}\log \left(\frac{\sqrt{x^2-2}-x}{\sqrt{2}}\right)\Bigg]+\cdots\\
&=&\frac{c}{2}\Bigg[- x\sqrt{x^2-2}-2\log \left(\sqrt{x^2-2}-x\right)+\log (2)\Bigg]+\cdots\,,
\end{eqnarray*}
from which we can directly read the left rate function for the smallest eigenvalue, viz
\begin{equation*}
\Psi_{-}(x)=\frac{1}{2} \Bigg[\log (2)- x\sqrt{x^2-2}-2\log \left(\sqrt{x^2-2}-x\right)\Bigg]\,,
\end{equation*}
in agreement with \cite{Majumdar2009}.
\section{Typical fluctuations around the point $(x,c_{\star}(x))$ for bulk eigenvalues}
\label{sec:tfa}
We focus now on the typical statistics for bulk eigenvalues, that is, those eigenvalues within the region $x\in(-\sqrt{2},\sqrt{2})$. From a mathematical viewpoint, this implies  doing an expansion of the rate function around the line $c_{\star}(x)$. As we are close to $c_{\star}(x)$ we can proceed by doing an expansion of the expression \eref{ratere} around $\alpha=0$. Treating the roots perturbatively we obtain the following behaviour:
\begin{eqnarray*}
\lambda_{-}&=&-\sqrt{2}-\frac{2\alpha}{-4-2\sqrt{2}x_\star}+\cdots\,,\\
\lambda_0&=&x_\star+(x-x_\star)-\frac{2\alpha}{2-x_\star^2}+\cdots\,,\\
\lambda_{+}&=&\sqrt{2}-\frac{2\alpha}{-4+2\sqrt{2}x_\star}+\cdots\,.
\end{eqnarray*}
We then proceed to perform this expansion to the various factors appearing in the rate function.\\
Starting with the fraction $c$  of eigenvalues to the left of $x$ one finds that
\begin{eqnarray}
 c&&=c_\star(x_\star)-\frac{\sqrt{2-x_\star^2}}{\pi}\delta x-\frac{2\alpha}{\pi\sqrt{2-x_\star^2}}\nonumber\\
 &&+\frac{\alpha}{\pi\sqrt{2-x^2_\star}}\log\left|\frac{1}{\sqrt{2}(2-x_\star^2)}\left(\delta x+\frac{2\alpha}{2-x_\star^2}\right)\right|+\cdots
\label{eq:expc}
\end{eqnarray}
with $\delta x=x_\star-x$. Expression \eref{eq:expc} can be inverted perturbatively, viz.
\begin{equation}
\alpha=-\frac{\pi\sqrt{2-x_\star^2}\delta c+(2-x_\star^2)\delta x}{2-\log\left|\frac{1}{\sqrt{2}}\left(\frac{\delta x}{2-x_\star^2}+\frac{2\pi\delta c}{(2-x_\star^2)^{3/2}}\right)\right|}\,,
\label{eq:expc2}
\end{equation}
with $\delta c=c-c_\star$.\\
Similarly for the other terms appearing in \eref{ratere} we find the following expansions:
\begin{equation*}
2\int_{\lambda_0}^{x} dz S_{-}(z)+\lambda_0^2-x^2=-\frac{2\pi \alpha}{\sqrt{2-x_\star^2}}+\cdots\,,
\end{equation*}
and
\begin{equation*}
\hspace{-2.5cm}\frac{\lambda_{+}^2}{2}-\log(\lambda_{+})+\int_{\lambda_{+}}^\infty dz\left(S_{-}(z)-\frac{1}{z}\right)=\frac{1+\log(2)}{2}+\frac{\alpha}{\sqrt{2-x^2_\star}}\left(\pi-\arccos\left(\frac{x_\star}{\sqrt{2}}\right)\right)+\cdots\,.
\end{equation*}
 Gathering all these results back into eq. \eref{ratere} we obtain the following expression:
\begin{equation*}
\hspace{-2.5cm}\Psi(c,x)=\frac{\alpha}{2}\Bigg[-\frac{1}{2}\delta x +\frac{1}{\sqrt{2-x^2_\star}}\Bigg(-\pi c+\frac{1}{2}x_\star\sqrt{2-x_\star^2}+\pi-\arccos\left(\frac{x_\star}{\sqrt{2}}\right)\Bigg)+\cdots\Bigg]\,,
\end{equation*}
and recalling that $2\pi c_\star=2\pi +x_\star\sqrt{2-x_\star^2}-2\arccos\left(\frac{x_\star}{\sqrt{2}}\right)$ we finally have:
\begin{equation}
\Psi(c_{\star}+\delta c,x_\star-\delta x)=-\frac{\alpha}{2}\left[\frac{1}{2} \delta x +\frac{\pi}{\sqrt{2-x^2_\star}}\delta c+\cdots\right]\,,
\label{eq:expR}
\end{equation}
 where  $\alpha=\alpha(\delta c,\delta x)$ is given by \eref{eq:expc2}.\\
At a given point $(x_\star,c_{\star})$ and if we want to look at either the typical fluctuations along the $x$-direction or the typical fluctuations along the $c$-direction, we must take either $\delta c=0$ or $\delta x=0$, respectively. This produces the following two expansions of the rate function:
\begin{eqnarray*}
\Psi(c_{\star},x_\star-\delta x)&=&\frac{(2-x_\star^2)}{4} \frac{(\delta x)^2}{2-\log\left|\frac{1}{\sqrt{2}}\left(\frac{\delta x}{2-x_\star^2}\right)\right|}\,,\\
\Psi(c_{\star}+\delta c,x_\star)&=&\frac{\pi^2}{2}\frac{(\delta c)^2}{2-\log\left|\frac{\sqrt{2}\pi\delta c}{(2-x_\star^2)^{3/2}}\right|}\,.
\end{eqnarray*}
\begin{figure*}[t]
\includegraphics[width=7cm,height=5cm]{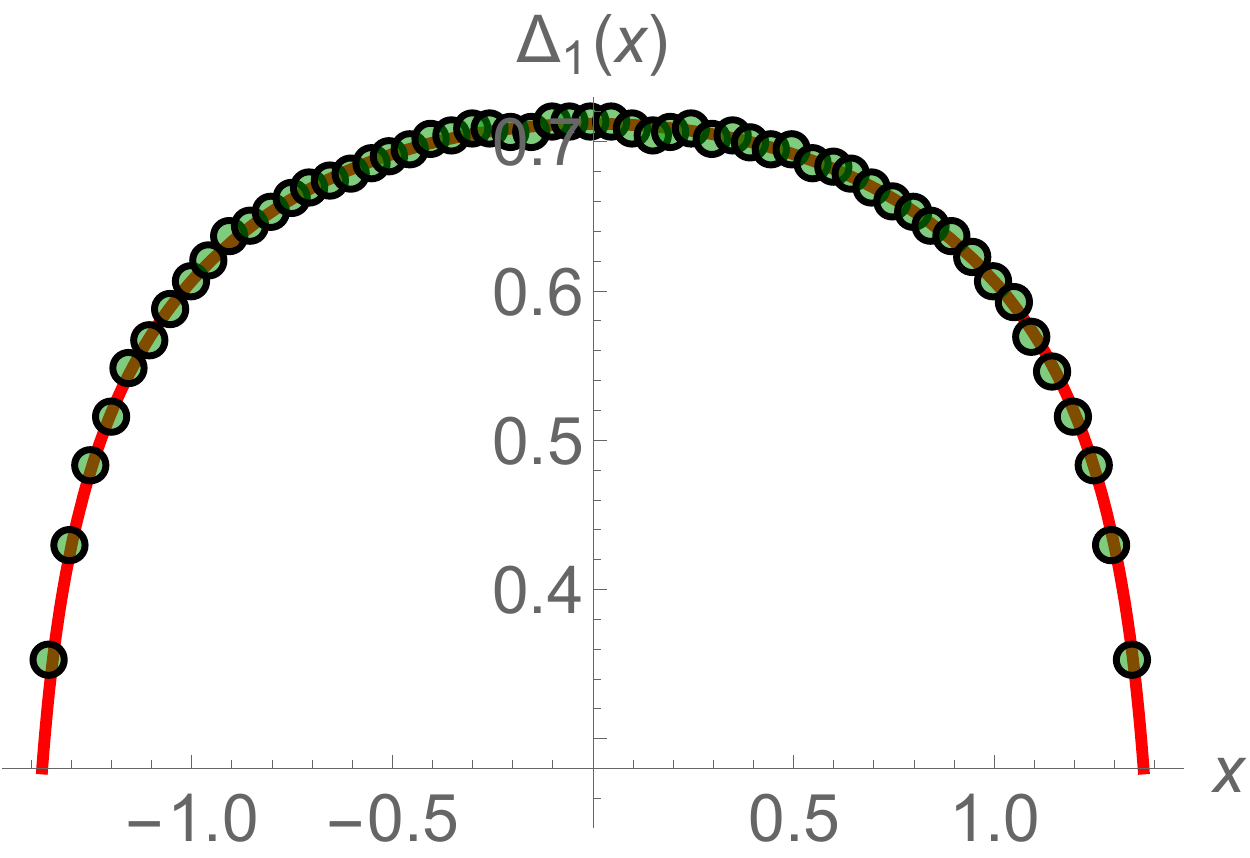}\quad\quad\quad\includegraphics[width=7cm,height=5cm]{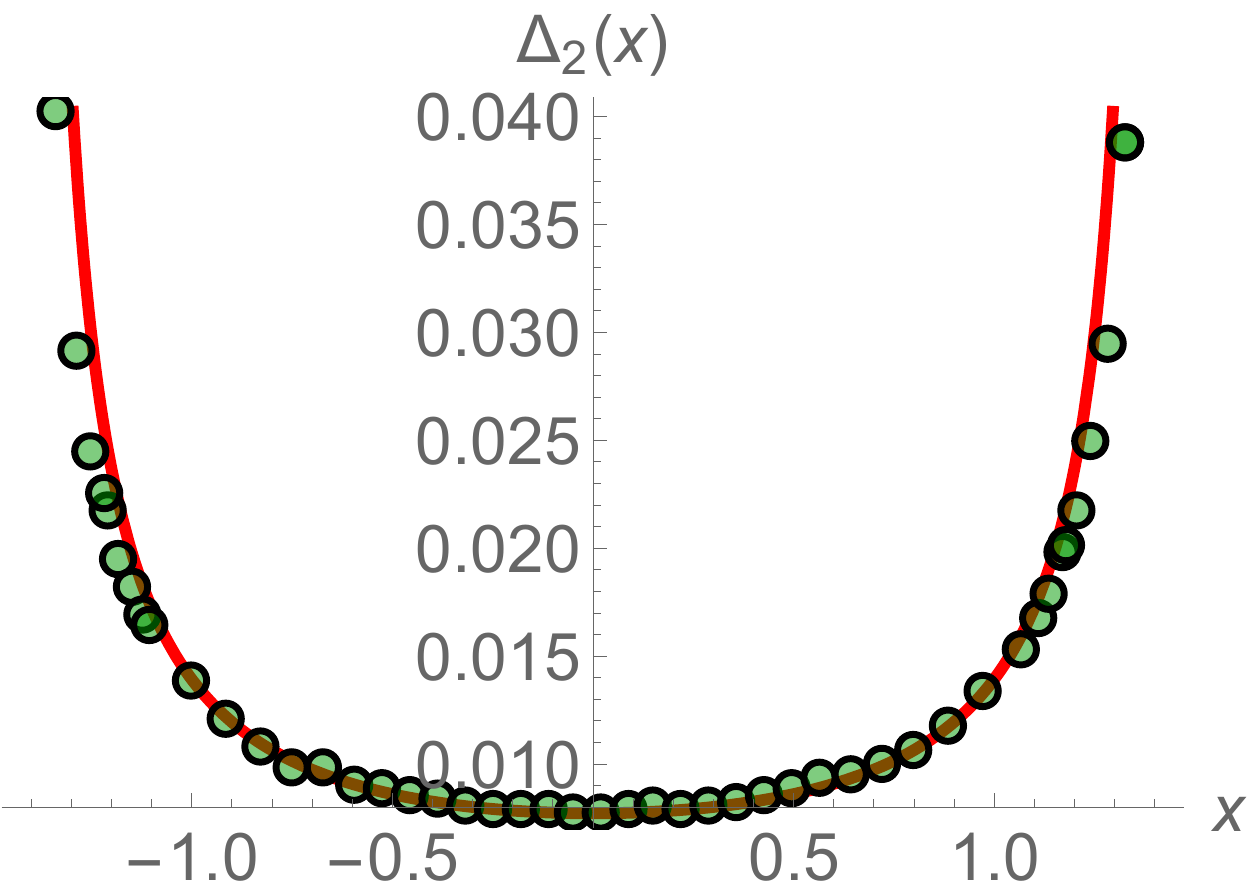}
\caption{Comparison between Monte Carlo simulations and Gustavsson's formulas. For $\Delta_2(x)$ we have taken $N=500$ eigenvalues and $\beta=1$. We have relaxed the Coulomb fluid using Metropolis for $5\cdot 10^5$ Monte Carlo steps. After that we have used a window of $1\cdot 10^6$ Monte Carlo steps to sample the mean value and variance of each eigenvalue $y_i$ with $y_{1}<\cdots<y_N$ each 100 steps, resulting in the reported figure. For $\Delta_1(x)$ we have taken $N=100$ eigenvalues and $\beta=1$.  We have relaxed the Coulomb gas for $10^6$ Monte Carlo steps and we have used a window of $10^7$ to take samples of $N_{x}$ every 100 Monte Carlo steps. We repeat this process for various values $x$. In both cases we have added a small constant to have the theoretical results on top of the numerical ones.}
\label{GVar2}
\end{figure*}
Recalling that for large $N$ we have that $F_{X}(x)\sim f_{X}(x)$ and  assuming, moreover, that that the fluctuations are symmetric around $(c_{\star},x_\star)$ we  find  the following results for the Gaussian fluctuations of $y_k$ and $\mathcal{N}_{x}$
\begin{eqnarray*}
\hspace{-1.5cm} f_{Y_{k}}(y_{k})\sim \exp\left[-\frac{(y_k-x_\star \sqrt{N})^2}{2\Delta_2}\right]\,,\quad\quad f_{\mathcal{N}_x}(n_x)\sim \exp\left[-\frac{(n_x-c_\star N)^2}{2\Delta_1}\right]\,,
\end{eqnarray*}
with variances $\Delta_1=\frac{\log [(2-x_\star^2)^{3/2}N]}{\beta \pi^2}$ and $\Delta_{2}=\frac{\log N}{\beta N(2-x_\star^2)}$. Here we have used that  $\delta c=c-c_\star=(n_x-c_\star N)/N$ and $\delta x=x_\star-x=(x_\star\sqrt{N}-y_k)/\sqrt{N}$. The expressions for $\Delta_1$ and $\Delta_2$ are in agreement with those found in \cite{Gustavsson2005,Rourke2010}.  In Fig. \ref{GVar2} we compare the analytical results of $\Delta_1$ and $\Delta_2$ with standard Monte Carlo simulations of  the Coulomb fluid.
\section{Comparison with Monte Carlo simulations}
\label{sec:mcs}
To check the validity of all our derivations we have performed thorough Monte Carlo simulations. Let us first recall that given the jPDF of eigenvalues of the Gaussian ensemble
\begin{equation}
P(\bm{y})=\frac{1}{A_0}e^{-\frac{\beta}{2}F(\bm{y})}\,,\quad F(\bm{y})=\sum_{i=1}^Ny_i^2-\sum_{i\neq j}\log|y_i-y_j|\,,
\label{eq:MC}
\end{equation}
the standard Metropolis algorithm consists in the following steps: i) pick a $k\in\{1,\ldots,N\}$ at random; ii) propose an update $y_{k}\to y_{k}'$; iii) calculate the energy difference $\Delta F$ due to the change; iv) accept the change with rate $p=\min\left\{1,e^{-\frac{\beta}{2}\Delta F}\right\}$. The energy difference due to this move is
\begin{equation*}
\Delta F=(y_k')^2-y_k^2-2\sum_{j(\neq k)}^N\log\left|\frac{y_k'-y_j}{y_k-y_j}\right|\,.
\end{equation*}
In what follows, we explain how we modify the standard Metropolis algorithm to improve the performance when comparing with our analytical results.
\subsection{Monte Carlo simulations for the constrained density $\rho(\lambda)$ and for the action $S_0(c,x)$}
As already noted in Figs. \ref{rho} and \ref{rate} we have compared our analytical findings with Monte Carlo Metropolis simulations using a slightly modified Metropolis algorithm. Given a position $x$ of the barrier, to simulate a given fraction $c$, we choose a total number of eigenvalues $N$ and divide them into those to the left of the barrier, $N_{{\rm left}}$, and those to the right of the barrier, $N_{{\rm right}}$. Initial conditions are chosen so that $c=N_{{\rm left}}/N$. Then one performs the standard Metropolis algorithm, but the updates are  chosen so that $N_{{\rm left}}$ remains constant throughout the Monte Carlo updating. This can  be done without adding extra rejection to the Metropolis algorithm. After letting the system thermalise, one performs averages over the Monte Carlo Markov chain generated by the algorithm.\\
The estimation of the density $\rho(\lambda)$ by Monte Carlo, which appears in Fig. \ref{rho}, is done straightforwardly and does not required much thought. The estimation of the action (or equivalently the rate function $\Psi(c,x)$ as shown in Fig. \ref{rate}) requires a bit of care, though. Here we must remember that when doing the analytics we have gone from $y_{i}\to\lambda_{i}$ and we have ignored constant terms which depend on $N$. Tracing back and correcting for those factors, we find that the action is to be estimated by the formula
\begin{equation*}
S_0(c,x)=\frac{1}{N^2}\left[\bracket{F(\bm{y})}_{{\rm thermal}}+\frac{1}{2}N(N-1)\log N\right]\,,
\end{equation*}
with $F(\bm{y})$ defined in \eref{eq:MC} and where $\bracket{\cdots}_{{\rm thermal}}$ stands for the thermal averaging, which is estimated by averaging over the Monte Carlo Markov chain generated by the algorithm.
\subsection{Modified Metropolis algorithm for the statistics of $\lambda_k$}
\begin{figure}
\centering
\includegraphics[width=8.5cm,height=6cm]{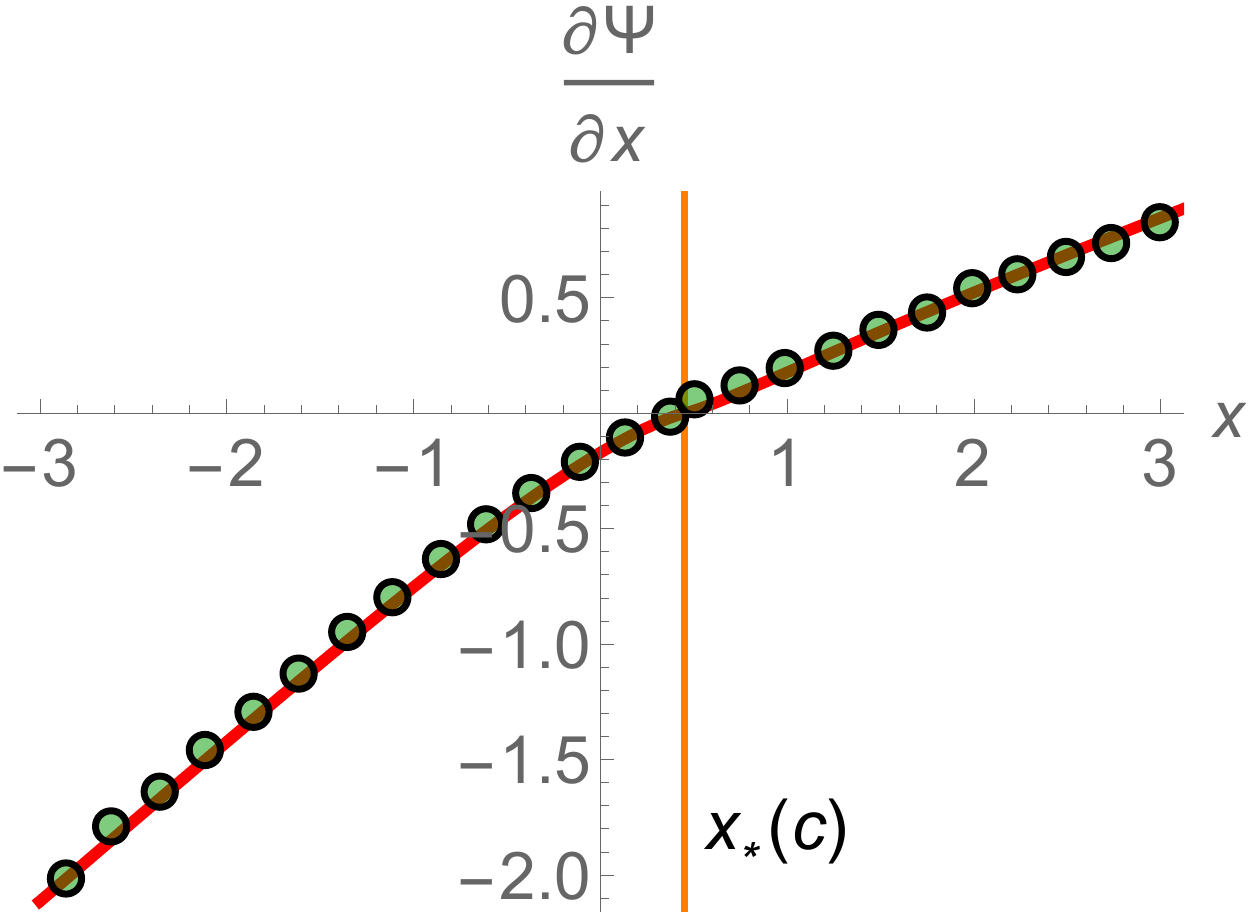}
\caption{Comparison between Monte Carlo simulations (solid green circles) and analytical result for $\partial_{x}\Psi(c,x)$. For the simulations, we have used $N=20$, $N_{{\rm left}}=14$ (so that $c=7/10$), and $\beta=1$. To estimate each point in the figure, the Coulomb gas is first relaxed over $5\cdot10^6$ ($5\cdot10^7$ for $x>x_{\star}(c)$) Monte Carlo steps. Then throughtout a window of $5\cdot10^ 7$ Monte Carlo steps, one records the statistics of the $k$-th eigenvalue (the one closer to of barrier, to the left if $x<x_{\star}$ or to the right if $x>x_{\star}$) every 100 Monte Carlo steps. The $5\cdot10^5$ records are used to construct a histogram of the cdf of the $k$-th eigenvalue, using 20 bins. Finally a 5-point formula is used to estimate the derivative of the logaritm of the cumulative.}
\label{MM}
\end{figure}
We finish this part by briefly explaining how we estimate the rate function for bulk eigenvalues. As it has been explained somewhere else, if we were to use the standard Metropolis algorithm to construct an histogram for ${\rm Prob}[\lambda_k<x]$ it would require a prohibitively large number of samples to obtain a reliable statistics for $x$ away from its typical value $x_\star$. However, this can be written as ${\rm Prob}[\lambda_k<x| x\leq y]K_{y}$, for a constant $K_y$. Essentially, one is putting a barrier at $y$ so that $\lambda_{k}$ would be concentrated around $x\in[y-\delta,\delta]$. This gives directly the derivative of the rate function with respect to $x$,
\begin{equation*}
\Psi_{,x}(c,x)=-\frac{1}{\beta N^2} \frac{d}{dx}\log {\rm Prob}[\lambda_k<x| x\leq y]\,.
\end{equation*}
This nice little trick gives the branch of $\Psi(c,x)$ for $x<x_{\star}(c)$. A similar argument gives the branch corresponding to $x>x_{\star}(c)$.\\
The result of this algorithm is presented in Fig. \ref{MM}. Here we have taken $N=20$ with $N_{{\rm left}}=14$, which means that we are estimating the statistics of the 14-th eigenvalue or, for very large matrices, that eigenvalue with typical value $x_{\star}(c)$ corresponding to $c=7/10$. As we can see the results are strikingly good.
\section{Summary and Future work}
\label{sec:cfw}
In this work we have presented a method to obtain the full order statistics of the eigenvalues of the Gaussian ensemble. This has been achieved by introducing a rate function $\Psi(c,x)$ depending on two parameters $c$ and $x$. When $c=k/N$  the rate function $\Psi(k/N,x)$ gives the large deviations of the $k$-th eigenvalue as a function of $x$. Importantly, when $k=1$ (or $k=N$) the rate function $\Psi(k/N,x)$, when analysed carefully, provides both the fluctuations to the left and to the right of the typical value $-\sqrt{2}$ of $\lambda_1$. If we fix $x$, then $\Psi(c,x)$ provides the large deviations of the SIN, that is, the large deviations of the fraction $c$ of eigenvalues to the left of $x$, hence generalising the work on the index number problem.\\
It would be interesting to see how the method presented here can be  extended to other ensembles (e.g. Wishart, Jacobi, Cauchy), as it might  be that for some ensembles it is not possible to obtain exact expressions for the integrals involving the resolvent in terms of e.g. elliptic integrals. However, perturbations around the typical behaviour would be  certainly possible. The latter will help to see how the variances of typical fluctuations are affected when considering eigenvalues in a restricted subset of $\mathbb{R}$.  An analysis of the Coulomb fluid on a general external polynomial potential $V(x)$ would be of some interest too.\\
It will be interesting and challenging to see the connection with the known results using determinantal methods (and its relation to Painvel\'e systems)  applied to the $k$-th eigenvalue and valid for finite $N$.\\
Finally, the work presented here could also be extended to study  the order statistics of a pair of eigenvalues $\lambda_{k}$ and $\lambda_\ell$, with applications, for instance, to the statistics of the gap between consecutive eigenvalues or to the condition number. These are current lines of research.

\ack
The author acknowledges discussions with Pierpaolo Vivo and Fabio Cunden.  Part of this work has been funded by the program UNAM-DGAPA-PAPIIT IA101815.


\appendix
\section{Deriving $c(\alpha,x)$}
\label{app:ca}
In this section we derive an exact expression for the fraction $c$ as a function of $\alpha$ and $x$ in terms of complete elliptic integrals. To do so, we need to distinguish between the cases $c>c_{\star}(x)$ and $c<c_{\star}(x)$.
\subsection{Case $c>c_{\star}$}
Here the fraction $c$ of eigenvalues to the left of the barrier is given by
\begin{equation}
c(\alpha,x)=\frac{1}{\pi}\int_{\lambda_{-}}^x d\lambda\sqrt{\frac{(\lambda_{+}-\lambda)(\lambda_{0}-\lambda)(\lambda-\lambda_{-})}{x-\lambda}}\,,
\label{eq:ca}
\end{equation}
which is related to the the following integral
\begin{equation*}
I=\int_{d}^{y}\sqrt{\frac{(a-t)(b-t)(t-d)}{(c-t)}}\,,
\end{equation*}
with $\lambda_{+}=a$, $\lambda_{0}=b$, $x=c$, $\lambda_{-}=d$, and $y=c$.  This can be related to elliptic integrals (given by \cite{Byrd1971}, formula 252.36):
\begin{eqnarray*}
\hspace{-1.5cm}I&=&(a-d)^2(d-b)\tilde{\alpha}^2 g\int_0^{u_1}\frac{sn^2(u)dn^2(u)du}{(1-\tilde{\alpha}^2sn^2(u))^3}\\
\hspace{-1.5cm}g&=&\frac{2}{\sqrt{(a-c)(b-d)}}\,,\quad k^2=\frac{(a-b)(c-d)}{(a-c)(b-d)}\,,\tilde{\alpha}^2=\frac{d-c}{a-c}<0\,,\quad u_1=K(k)\,,
\end{eqnarray*}
with the following definition of the $V_{m}=\int_0^{u_1}\frac{du}{(1-\alpha^2 sn^2(u))^m}$  (formulas 336 at \cite{Byrd1971}). Gathering terms together and after massaging a bit we obtain the final expression:
\begin{eqnarray*}
\hspace{-1.2cm}c(\alpha,x)&=&\frac{1}{\pi}\frac{1}{4 \sqrt{(a-c) (b-d)}}\Bigg[(a-d) \left(a^2-2 d (a+b-c)\right.\\
\hspace{-1.2cm}&&\left.-2 a b+2 a c+b^2+2 b c-3 c^2+d^2\right) \Pi \left(\frac{d-c}{a-c},k\right)\\
 \hspace{-1.2cm}&-&(a-d) (a-c) (a-3 b+3 c-d) K\left(k\right)\\
 \hspace{-1.2cm}&& -(a-c) (b-d) (a+b-3 c+d) E\left(k\right)\Bigg]\,,\quad k^2=\frac{(a-b)(c-d)}{(a-c)(b-d)}\,.
\end{eqnarray*}
This  is still valid for general values of $a$, $b$, $c$, and $d$ (as soon as they are ordered as our roots). In our  case we have, however, that $a=\lambda_+$, $b=\lambda_{0}$, $c=x$ and $d=\lambda_{-}$. This implies that, the  coefficients $a$, $b$, $c$, and $d$ obey certain equalities, which allows us to  simplify further this finding, as presented in the text.
\subsection{Case $c<c_{\star}(x)$}
In this case the fraction $c$ of eigenvalues reads:
\begin{equation}
c(\alpha,x)=\frac{1}{\pi}\int_{\lambda_{-}}^{\lambda_0} d\lambda\sqrt{\frac{(\lambda_{+}-\lambda)(\lambda_{0}-\lambda)(\lambda-\lambda_{-})}{x-\lambda}}\,,
\label{eq:cb}
\end{equation}
which is related to the elliptic integral
\begin{equation*}
I=\int_{y}^{c}\sqrt{\frac{(a-t)(c-t)(t-d)}{b-t}}
\end{equation*}
with $a=\lambda_+$, $b=x$, $c=\lambda_0$, $d=\lambda_-$, and with $y=d$. This corresponds to the elliptic integral 253.34 page 111 in \cite{Byrd1971}, viz.
\begin{eqnarray*}
I&=(a-c)(b-c)(c-d)\tilde{\alpha}^2 g\int_0^{u_1}\frac{sn^2(u)cn^2(u)dn^2(u)}{(1-\tilde{\alpha}^2sn^2(u))^3}du\,,\\
k^2&=\frac{(a-b)(c-d)}{(a-c)(b-d)}\,,\quad g=\frac{2}{\sqrt{(a-c)(b-d)}}\,,\\
\tilde{\alpha}^2&=\frac{c-d}{b-d}\,,\quad u_1=K(k)\,.
\end{eqnarray*}
The final expression for fraction $c$ of eigenvalues becomes
\begin{eqnarray*}
\hspace{-1.2cm}c(\alpha,x)&=&\frac{1}{4\pi \sqrt{(a-c) (b-d)}}\Bigg[ (b-c)\left(a^2+2 a (b-c-d)\right.\\
\hspace{-1.2cm}&&\left.-3 b^2+2 b (c+d)+(c-d)^2\right) \Pi \left(\frac{c-d}{b-d},k\right)\\
\hspace{-1.2cm}&+&(b-c)(b-d) (-5 a+3 b+c+d) K\left(k\right)\\
\hspace{-1.2cm}&&-(a-c) (b-d) (a-3 b+c+d) E\left(k\right)\Bigg]\,,\quad k^2=\frac{(a-b)(c-d)}{(a-c)(b-d)}\,,
\end{eqnarray*}
which again can be simplified, as given in the main text recalling that the roots and the barrier $x$ obey certain equalities.
\section{Exact expressions for the integrals in Eq. \eref{ratere}}
\label{app:ei}
Let us find exact expressions for the integrals appearing in the expressions of the constants $A_1$ and $A_2$. Again we have  to make a distinction between cases $c>c_{\star}(x)$ and $c<c_{\star}(x)$.
\subsection{Case $c>c_{\star}(x)$}
\subsubsection{Integral $\int_{x}^{\lambda_0} dz S_{+}(z)$}
To evaluate the integral $\int_{x}^{\lambda_0} dz S_{+}(z)$, we note that it is related to the following  elliptic integral (255.39 page 120 from \cite{Byrd1971})
\begin{eqnarray*}
I_1&=&\int_{c}^{b} dt\sqrt{\frac{(t-a)(t-b)(t-d)}{t-c}}\\
&=&(a-b)^2(b-d)\frac{g}{\talpha^2}[-k^2\Pi(\talpha^2,k)+(2k^2-\talpha^2)V_2+(\talpha^2-k^2)V_3]\,,\\
k^2&=&\frac{(b-c)(a-d)}{(a-c)(b-d)}\,,\quad g=\frac{2}{\sqrt{(a-c)(b-d)}}\,,\quad \talpha^2=\frac{b-c}{a-c}\,,
\end{eqnarray*}
and with the already given the definitions for the $V$ functions. Gathering all these results and simplifying, it becomes:
\begin{eqnarray*}
I_1(a,b,c,d)&=&\frac{1}{4 \sqrt{(a-c) (b-d)}}\Bigg[(a-b)\left(a^2-2 d (a+b-c)\right.\\
&&\left.-2 a b+2 a c+b^2+2 b c-3 c^2+d^2\right) \Pi \left(\frac{b-c}{a-c},k\right)\\
&&-(a-b)(a-c) (a-b+3 c-3 d) K\left(k\right)\\
&&+(a-c) (b-d) (a+b-3 c+d) E\left(k\right)\Bigg]\,.
\end{eqnarray*}
Applied to our particular case yields:
\begin{equation*}
\int d\lambda' \rho(\lambda')\log\left|\frac{\lambda_0-\lambda'}{x-\lambda'}\right|=\frac{\lambda_0^{2}-x^2}{2}+I_1(\lambda_{+},\lambda_0,x,\lambda_-)\,.
\end{equation*}
As mentioned before, due the the roots obeying certain equalities, $I_1$ is simplified further as reported in the main text.
\subsubsection{Evaluation of $\int_{\lambda_{+}}^{\infty} dz \left[S_{-}(z)-\frac{1}{z}\right]$.}
Next, we move to evaluate $\int_{\lambda_{+}}^{\infty} dz \left[S_{-}(z)-\frac{1}{z}\right]$. Firstly, we notice that since the resolvent behaves at infinity as $S_{-}(z)=\frac{1}{z}+\mathcal{O}(z^{-2})$, the integral is clearly convergent. Here, to obtain an exact expression in terms of elliptic integrals we must work a bit harder. Essentially, we want to do the following integral and then evaluate the limit
\begin{eqnarray*}
\hspace{-1.5cm}I_2(a,b,c,d)&=&\lim_{y\to\infty}\int_{a}^{y}dt\Bigg[\sqrt{\frac{(t-a)(t-b)(t-d)}{(t-c)}}-\frac{1}{2} (-a-b+c-d)-t\\
\hspace{-1.5cm}&&-\frac{-a^2+2 a (b-c+d)-b^2+2 b (d-c)+(c-d) (3 c+d)}{8 t}\Bigg]\,,
\end{eqnarray*}
where we have subtracted the divergences at infinity. Or in other words: we must evaluate the integral and study its asymptotic behaviour for $y\to\infty$, viz.
\begin{equation*}
\hspace{-1.5cm}\int_{a}^{y}dt\sqrt{\frac{(t-a)(t-b)(t-d)}{(t-c)}}\sim Ay^2+By+C\log(y)+D+\mathcal{O}(y^{-1})\,.
\end{equation*}
From formula 258.36 page 132 in \cite{Byrd1971} we arrive at\footnote{There is a typo of this formula in \cite{Byrd1971}. The formula reported here is the correct one.}
\begin{eqnarray*}
I_3&=&\int_{a}^{y}dt\sqrt{\frac{(t-a)(t-b)(t-d)}{(t-c)}}\\
&=&(a-b)^2(a-d)\frac{g}{\talpha^2}\left[-\Pi(u_1,\talpha^2,k)+(2-\talpha^2)V_2(u_1)+(\talpha^2-1)V_3(u_1)\right]\,,\\
k^2&=&\frac{(b-c)(a-d)}{(a-c)(b-d)}\,,\quad g=\frac{2}{\sqrt{(a-c)(b-d)}}\,,\\
sn(u_1)&=&\sin\varphi\,,\quad\varphi=\sin^{-1}\sqrt{\frac{(b-d)(y-a)}{(a-d)(y-b)}}\,,\\
\talpha^2&=&\frac{a-d}{b-d}\,,
\end{eqnarray*}
with the following definitions for the $V$ functions:
\begin{eqnarray*}
V_0&=&u=F(\varphi,k)\,,\\
V_{1}&=&\Pi(\varphi,\tilde{\alpha}^2,k)\,,\\
V_{2}&=&\frac{1}{2(\tilde{\alpha}^2-1)(k^2-\tilde{\alpha}^2)}\Bigg[\tilde{\alpha}^2E(u_1,k)+(k^2-\tilde{\alpha}^2)u_1\\
&&+(2\tilde{\alpha}^2k^2+2\tilde{\alpha}^2-\tilde{\alpha}^4-3k^2)\Pi(\varphi,\tilde{\alpha}^2,k)-\frac{\talpha^4sn(u_1)cn(u_1)dn(u_1)}{1-\talpha^2sn^2(u_1)}\Bigg]\,,\\
V_3&=&\frac{1}{4(1-\tilde{\alpha}^2)(k^2-\tilde{\alpha}^2)}\Bigg[k^2 V_0+2(\tilde{\alpha}^2k^2+\tilde{\alpha}^2-3k^2)V_1\\
&&+3(\tilde{\alpha}^4-2\tilde{\alpha}^2k^2-2\tilde{\alpha}^2+3k^2)V_2+\frac{\talpha^4sn(u_1)cn(u_1)dn(u_1)}{(1-\talpha^2sn^2(u_1))^2}\Bigg]\,.
\end{eqnarray*}
In the limit $y\to\infty$ it is possible to see that the divergences are coming from the following terms:
\begin{eqnarray*}
\hspace{-2cm}\Pi\left(\sin^{-1}\sqrt{\frac{(b-d)(y-a)}{(a-d)(y-b)}},\frac{a-d}{b-d},k\right)&&=-\Pi\left(\sin^{-1}\sqrt{\frac{(b-d)}{(a-d)}},\frac{b-c}{a-c},k\right)\\
\hspace{-2cm}&&+F\left(\sin^{-1}\sqrt{\frac{(b-d)}{(a-d)}}, k\right)+\mathcal{E}_1(a,b,c,d)\\
\hspace{-2cm}&&+A\log(y)+\cdots\,,\\
\hspace{-2cm}\frac{\talpha^4sn(u_1)cn(u_1)dn(u_1)}{1-\talpha^2sn^2(u_1)}&&=\mathcal{E}_2(a,b,c,d)+A y+B\frac{1}{y}+\mathcal{O}(y^{-2})\,,\\
\hspace{-2cm}\frac{\talpha^4sn(u_1)cn(u_1)dn(u_1)}{(1-\talpha^2sn^2(u_1))^2}&&=\mathcal{E}_3(a,b,c,d)+Ay^2+By+Cy^{-1}+\mathcal{O}(y^{-2})\,,
\end{eqnarray*}
where we have defined
\begin{eqnarray*}
\mathcal{E}_1(a,b,c,d)&=&\frac{\sqrt{(a-c) (b-d)} }{2 (a-b)}\log \left(\frac{4}{a+b-c-d}\right)\,,\\
\mathcal{E}_2(a,b,c,d)&=&-\frac{(a-d) (a-b+c+d)}{2 \sqrt{a-c} (b-d)^{3/2}}\,,\\
\mathcal{E}_3(a,b,c,d)&=&\frac{(a-d) \left(a^2-2 a (b+c+d)+b^2-2 b (c+d)+(c-d)^2\right)}{8 (b-a) \sqrt{a-c} (b-d)^{3/2}}\,.
\end{eqnarray*}
This allows us to obtain the following expression for $I_2$
\begin{eqnarray*}
\hspace{-2cm}I_2(a,b,c,d)&=&\frac{(a-b) \left(a^2-2 a (b-c+d)+b^2+2 b c-2 b d-3 c^2+2 c d+d^2\right) \Pi \left(\theta,\frac{b-c}{a-c},k\right)}{4 \sqrt{(a-c) (b-d)}}\\
\hspace{-2cm}&&-\frac{(a-b) (a-c) (a-b+3 c-3 d) F(\theta ,k)}{4 \sqrt{(a-c) (b-d)}}\\
\hspace{-2cm}&&+\frac{1}{4} \sqrt{(a-c) (b-d)} (a+b-3 c+d) E(\theta ,k)\\
\hspace{-2cm}&&+\frac{1}{16 }\Bigg[ -2 \big(a^2-2 a (b-c+d)+b^2\\
\hspace{-2cm}&&+2 b c-2 b d-3 c^2+2 c d+d^2\big) \log \left(\frac{4 a}{a+b-c-d}\right)\\
\hspace{-2cm}&&+a^2-2 d (a-b+c)-6 a b+6 a c-3 b^2+10 b c-7 c^2+d^2\Bigg]
\end{eqnarray*}
Recalling again that the coefficients are related to the roots, the final expression takes the form
\begin{eqnarray*}
\hspace{-2cm}I_2(a,b,c,d)&=&-\frac{(a-b) (a-c) (a+c-d)}{2 \sqrt{(a-c) (b-d)}} F(\theta ,k)-\frac{c}{2} \sqrt{(a-c) (b-d)} E(\theta,k)\\
\hspace{-2cm}&&+\frac{2 (a-b) }{\sqrt{(a-c) (b-d)}}\Pi \left(\theta,\frac{b-c}{a-c},k\right)+\frac{1}{2} \left(d(b-c) +1-2 \log \left(\frac{2 a}{-d}\right)\right)\,,\\
\hspace{-2cm}a&&=\lambda_{+}\,,\quad b=\lambda_{0}\,,\quad c=x\,,\quad d=\lambda_{-}\,,\quad k^2=\frac{(b-c)(a-d)}{(a-c)(b-d)}\,,\\
\hspace{-2cm}\theta&&=\sin^{-1}\sqrt{\frac{(b-d)}{(a-d)}}\,.
\end{eqnarray*}
Thus we have
\begin{equation*}
I_2(\lambda_{+},\lambda_{0},x,\lambda_{-})+\log\lambda_+=\int d\lambda \rho(\lambda)\log\left|\lambda_{+}-\lambda\right|\,.
\end{equation*}
\subsection{Case $c<c_{\star}(x)$}
\subsubsection{Evaluation of  $\int^{x}_{\lambda_0} dz S_{-}(z)$}
As before we can exploit the known results on elliptic integrals to obtain exact expressions of the integrals appearing in constants $A_1$ and $A_2$. Indeed,  the integral $\int^{x}_{\lambda_0} dz S(z)$, is given by formula 254.35 page 115 of \cite{Byrd1971}
\begin{eqnarray*}
J_1&=&\int_{c}^{b}dt\sqrt{\frac{(a-t)(t-c)(t-d)}{(b-t)}}\\
&=&(c-d)^2(a-c)\frac{g}{\talpha^2}[-k^2\Pi(\talpha^2,k)+(2k^2-\talpha^2)V_2+(\talpha^2-k^2)V_3]\,,\\
k^2&=&\frac{(b-c)(a-d)}{(a-c)(b-d)}\,,\quad g=\frac{2}{\sqrt{(a-c)(b-d)}}\,,\quad \talpha^2=\frac{b-c}{b-d},\quad u_1=K(k)\,,
\end{eqnarray*}
which Mathematica can help us to simplify as follows:
\begin{eqnarray*}
J_1(a,b,c,d)&=&\frac{1}{4\sqrt{(a-c)(b-d)}}\Bigg[ (c-d)\left(a^2+2 a (b-c-d)\right.\\
&&\left.-3 b^2+2 b (c+d)+(c-d)^2\right) \Pi \left(\frac{b-c}{b-d}|\frac{(b-c) (a-d)}{(a-c) (b-d)}\right)\\
&&+(c-d)(b-d) (-3 a+3 b-c+d) K\left(\frac{(b-c) (a-d)}{(a-c) (b-d)}\right)\\
&&-(a-c) (b-d) (a-3 b+c+d) E\left(\frac{(b-c) (a-d)}{(a-c) (b-d)}\right)\Bigg]\,.
\end{eqnarray*}
Thus we have that
\begin{equation*}
\int_{\lambda_0}^{x}dz S_{-}(z)=\int d\lambda'\rho(\lambda')\log\left|\frac{x-\lambda'}{\lambda_0-\lambda'}\right|=\frac{x^2-\lambda_0^2}{2}-J_{1}(\lambda_+,x,\lambda_{0},\lambda_{-})\,.
\end{equation*}
Notice that we can simplify the expression for $J_1(a,b,c,d)$ as presented in the main text, since the roots and $x$ obey certain equalities.
\subsubsection{evaluation of $\int_{\lambda_{+}}^\infty dz\left[S_{-}(z)-\frac{1}{z}\right]$}
The evaluation of  $\int_{\lambda_{+}}^\infty dz\left[S_{-}(z)-\frac{1}{z}\right]$ is straightforward. As the parameters seem to be the same as before we can readily say that the asymptotic solution for $J_2$ is
\begin{eqnarray*}
\hspace{-2cm}J_{2}(a,b,c,d)&=&\frac{(a-b) \left(a^2+2 a (b-c-d)-3 b^2+2 b (c+d)+(c-d)^2\right) }{4 \sqrt{(a-c) (b-d)}}\Pi \left(\theta,\frac{b-c}{a-c},k\right)\\
\hspace{-2cm}&&-\frac{(a-b) (a-c) (a+3 b-c-3 d)}{4 \sqrt{(a-c) (b-d)}} F(\theta ,k)\\
\hspace{-2cm}&&+\frac{1}{4} \sqrt{(a-c) (b-d)} (a-3 b+c+d) E(\theta,k)\\
\hspace{-2cm}&&+\frac{1}{16} \Bigg(2 \big(a^2+2 a (b-c-d)-3 b^2\\
\hspace{-2cm}&&+2 b (c+d)+(c-d)^2\big) \log \left(\frac{a+b-c-d}{4 a}\right)\\
\hspace{-2cm}&&+a^2-2 a (-b+c+d)+5 b^2-6 b (c+d)+c^2+6 c d+d^2\Bigg)\,,
\end{eqnarray*}
with further simplications  due to the relations of the coefficients, viz.
\begin{eqnarray*}
\hspace{-2cm}J_{2}(a,b,c,d)&=&-\frac{(a-b) (a-c) (a+b-d)}{2 \sqrt{(a-c) (b-d)}} F(\theta,k)-\frac{1}{2} b \sqrt{(a-c) (b-d)} E(\theta ,k)\\
\hspace{-2cm}&&+\frac{2 (a-b) }{ \sqrt{(a-c) (b-d)}}\Pi \left(\theta,\frac{b-c}{a-c},k\right)+\frac{1}{2} \left(-2 \log \left(2\right)+(b-c) (b-d)+1\right)\,,\\
\hspace{-2cm}a&&=\lambda_{+}\,,\quad b=x\,,\quad c=\lambda_{0}\,,\quad d=\lambda_{-}\,,\quad k^2=\frac{(b-c)(a-d)}{(a-c)(b-d)}\,,\\
\hspace{-2cm}\theta&&=\sin^{-1}\sqrt{\frac{(b-d)}{(a-d)}}\,.
\end{eqnarray*}
We can then write:
\begin{eqnarray*}
\int d\lambda \rho(\lambda)\log|\lambda_{+}-\lambda|&=&\log(\lambda_{+})-\int_{\lambda_{+}}^\infty dz\left(S_{-}(z)-\frac{1}{z}\right)\\
&=&J_2(\lambda_{+},x,\lambda_0,\lambda_{-})+\log(\lambda_+)\,.
\end{eqnarray*}

\end{document}